\documentclass[aps,showpacs,twocolumn,
superscriptaddress]{revtex4}
\usepackage{graphicx}
\usepackage{dcolumn}
\usepackage{bm}
\usepackage{color}
\usepackage[normalem]{ulem} 
\usepackage[dvipsnames]{xcolor} 
\usepackage{hyperref}
\usepackage{orcidlink}
 \usepackage{tikz}
\usetikzlibrary{arrows.meta,calc,positioning,decorations.pathmorphing}
\hypersetup{
  colorlinks=true,        
  linkcolor=blue,         
  citecolor=cyan,         
}
\usepackage{mathrsfs}
\usepackage[utf8]{inputenc}
\usepackage{mathtools}
\usepackage{doi}
\usepackage{amsmath}
\usepackage{amssymb}

\usepackage{scalerel}
\usepackage{tikz}
\usetikzlibrary{svg.path}

\definecolor{orcidlogocol}{HTML}{A6CE39}
\tikzset{
  orcidlogo/.pic={
    \fill[orcidlogocol] svg{M256,128c0,70.7-57.3,128-128,128C57.3,256,0,198.7,0,128C0,57.3,57.3,0,128,0C198.7,0,256,57.3,256,128z};
    \fill[white] svg{M86.3,186.2H70.9V79.1h15.4v48.4V186.2z}
                 svg{M108.9,79.1h41.6c39.6,0,57,28.3,57,53.6c0,27.5-21.5,53.6-56.8,53.6h-41.8V79.1z M124.3,172.4h24.5c34.9,0,42.9-26.5,42.9-39.7c0-21.5-13.7-39.7-43.7-39.7h-23.7V172.4z}
                 svg{M88.7,56.8c0,5.5-4.5,10.1-10.1,10.1c-5.6,0-10.1-4.6-10.1-10.1c0-5.6,4.5-10.1,10.1-10.1C84.2,46.7,88.7,51.3,88.7,56.8z};
  }
}

\newcommand\orcidicon[1]{\href{https://orcid.org/#1}{\mbox{\scalerel*{
\begin{tikzpicture}[yscale=-1,transform shape]
\pic{orcidlogo};
\end{tikzpicture}
}{|}}}}

\begin{document}

\title{Horizon-Brightened Acceleration Radiation and Optical Signatures of Generic Regular Black Holes from Nonlinear Electrodynamics}

\author{Uktamjon Uktamov\orcidlink{0009-0003-0423-2474}} \email{uktam.uktamov11@gmail.com}
\affiliation{School of Physics, Harbin Institute of Technology, Harbin 150001, People’s Republic of China}
\affiliation{University of Tashkent for Applied Sciences, Str. Gavhar 1, Tashkent 100149, Uzbekistan}
\affiliation{Tashkent State Technical University, Tashkent 100095, Uzbekistan}
\affiliation{Institute for Advanced Studies, New Uzbekistan University,
Movarounnahr str. 1, Tashkent 100000, Uzbekistan
.}

\author{Ali \"Ovg\"un \orcidlink{0000-0002-9889-342X}}
\email{ali.ovgun@emu.edu.tr}
\affiliation{Physics Department, Faculty of Arts and Sciences, Eastern Mediterranean University, Famagusta, 99628 North Cyprus via Mersin 10, Turkiye.}

\author{Reggie C. Pantig \orcidlink{0000-0002-3101-8591}} 
\email{rcpantig@mapua.edu.ph}
\affiliation{Physics Department, School of Foundational Studies and Education, Map\'ua University, 658 Muralla St., Intramuros, Manila 1002, Philippines.}

\author{Bobomurat Ahmedov \orcidlink{0000-0002-1232-610X}} 
\email{ahmedov@astrin.uz}
\affiliation{School of Physics, Harbin Institute of Technology, Harbin 150001, People’s Republic of China}
\affiliation{Institute for Advanced Studies, New Uzbekistan University,
Movarounnahr str. 1, Tashkent 100000, Uzbekistan
.}
\affiliation{Institute of Theoretical Physics, National University of Uzbekistan, Tashkent 100174, Uzbekistan.}

\date{\today}

\begin{abstract}
We investigate horizon–brightened acceleration radiation (HBAR) and optical signatures for a broad class of regular black holes sourced by nonlinear electrodynamics. The spacetimes considered are static, spherically symmetric, and nonsingular, and they include Bardeen-like, and Hayward-like regular black-hole limits as special cases. We characterize the horizon structure and thermodynamic properties, and we compute key optical observables by determining the photon-sphere location and the corresponding shadow size as seen by distant observers, including controlled perturbative limits and full numerical solutions. Using angular-size constraints for Sgr A* and M87* from the Event Horizon Telescope and the GRAVITY collaboration, we perform a Markov Chain Monte Carlo analysis to infer the admissible parameter ranges of the model and to quantify degeneracies among the black-hole mass and nonlinear-electrodynamic parameters. On the quantum side, we develop the near-horizon reduction relevant for HBAR, showing that the dominant sector governing the detector response exhibits conformal behavior and leads to a thermal excitation spectrum governed by the horizon temperature. We formulate a Lindblad master-equation description for the radiation field, identify the thermal steady state, and derive an HBAR entropy–energy relation consistent with a Clausius-type first law. Finally, we establish a Wien-type displacement law for the HBAR spectrum, expressing the peak wavelength in terms of horizon thermodynamic, thereby providing an additional observable link between nonlinear electrodynamics, regularity, and near-horizon quantum radiation.

\end{abstract}

\maketitle


\section{INTRODUCTION} \label{sec1}

General relativity provides a geometric description of gravitation in which spacetime curvature governs the motion of matter and radiation, while matter fields back-react through the gravitational field equations \cite{Einstein1916}. Black holes occupy a central position in this framework: they are both the most compact classical predictions of the theory and a laboratory where the interplay between geometry, quantum field theory, and thermodynamics becomes unavoidable \cite{Bardeen1973}. Semiclassical considerations reveal that horizons behave as thermodynamic systems, exhibiting an effective temperature and an entropy that scales with horizon area \cite{Bekenstein1973, Hawking1975}. These insights have reshaped our understanding of what a black hole is, yet they simultaneously sharpen the foundational question: what microscopic degrees of freedom account for horizon thermodynamics and how robust are the standard conclusions under changes in the underlying matter sector and short-distance physics \cite{Strominger1996, Ashtekar1998}.

A complementary motivation comes from the long-standing tension between classical gravitational collapse and the occurrence of curvature singularities \cite{Penrose1965, HawkingPenrose1970}. Although singularity theorems establish that singular behavior is generic under broad assumptions, it is widely expected that a consistent ultraviolet completion of gravity should resolve such pathologies. One pragmatic and widely studied route is to consider regular black holes, in which the geometry remains nonsingular everywhere while retaining an event horizon and an asymptotically well-behaved exterior. A particularly economical realization of this idea arises within general relativity coupled to nonlinear electrodynamics (NED) \cite{Hayward2006, AyonBeato1999, AyonBeato2000}. In that setting, the electromagnetic sector departs from Maxwell theory at high field strengths, supplying an effective stress-energy tensor capable of smoothing the central region without introducing ad hoc modifications to the gravitational action \cite{Bronnikov2001}. Such constructions provide a controlled arena in which one can test how horizon thermodynamics and quantum-field-theoretic processes depend on the detailed matter content that supports the geometry \cite{Toshmatov2019}.
At the same time, black holes are no longer purely theoretical. Horizon-scale observations and precision tests of strong-field gravity increasingly probe the near-horizon and photon-region structure through electromagnetic signatures \cite{EHT2019, EventHorizonTelescope:2022wkp}. Two diagnostics are especially relevant for spherically symmetric compact objects. First, the shadow size, controlled by the critical behavior of null trajectories in the photon region, offers a direct observable associated with the optical appearance of the object \cite{Synge1966, Falcke2000}. Second, weak-field gravitational lensing, encoded in the deflection angle for light rays passing far outside the horizon, provides a complementary probe of the asymptotic and intermediate-radius geometry \cite{Will2014}. Regular black holes supported by nonlinear electrodynamics can reproduce the standard far-field behavior while differing in the strong-field regime, so shadow and lensing calculations offer a concrete way to connect theoretical models with observational constraints, without presupposing that the central region is singular \cite{Abdujabbarov2016, Eiroa2011}. In parallel with these geometric and observational developments, there has been renewed interest in operational and microscopic viewpoints on horizon thermodynamics. A particularly instructive perspective treats horizon-associated radiation not only as an abstract property of quantum fields in curved spacetime, but as an emergent phenomenon that can be read out through matter–field interactions \cite{Unruh1976, Davies1975}.

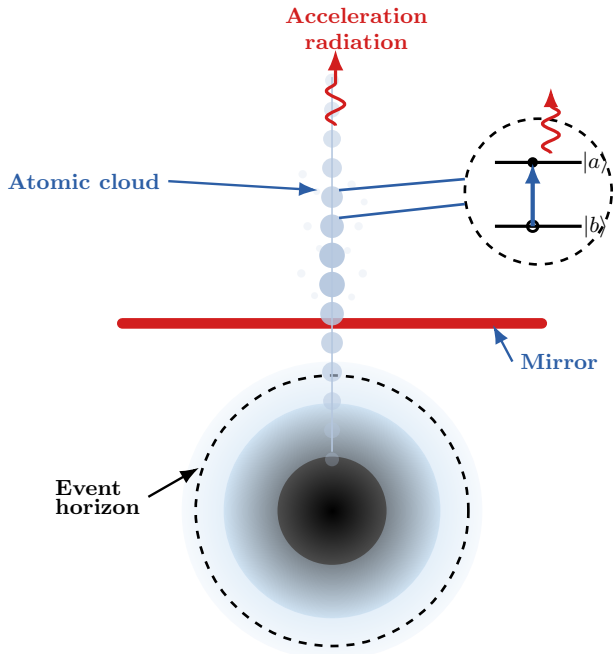
\begin{figure}[t]
\centering
\begin{tikzpicture}[
  >=Latex,
  font=\small,
  scale=0.92, transform shape 
]

\definecolor{HBARblue}{RGB}{44,94,165}
\definecolor{HBARlight}{RGB}{170,205,235}
\definecolor{HBARred}{RGB}{210,30,30}
\definecolor{HBARgray}{RGB}{125,125,125}

\coordinate (BH) at (0,0);

\shade[inner color=HBARlight!85, outer color=HBARlight!10] (BH) circle (2.15);
\shade[inner color=black!85, outer color=HBARlight!55] (BH) circle (1.55);
\shade[inner color=black, outer color=black!70] (BH) circle (0.78);

\draw[dashed, line width=1.0pt] (BH) circle (1.95);

\node[align=left] (EHlab) at (-3.35,0.20) {\bfseries Event\\[-1mm]\bfseries horizon};
\draw[-{Latex[length=2.5mm]}, line width=0.9pt] (EHlab.east) -- (-1.90,0.62);

\draw[line width=4.0pt, HBARred, line cap=round] (-3.0,2.70) -- (3.0,2.70);

\node[HBARblue, font=\bfseries] (Mlab) at (3.25,2.15) {Mirror};
\draw[-{Latex[length=2.5mm]}, HBARblue, line width=0.9pt] (Mlab.west) -- (2.30,2.68);

\draw[HBARblue!35, line width=0.6pt] (0,6.25) -- (0,0.85);

\foreach \k in {0,...,13}{
  \pgfmathsetmacro{\y}{6.20-0.42*\k}
  \pgfmathsetmacro{\amp}{max(0,1-abs(\k-6.5)/6.5)}
  \pgfmathsetmacro{\r}{0.10 + 0.09*\amp}
  \pgfmathsetmacro{\op}{0.30 + 0.70*\amp}
  \fill[HBARblue!35, opacity=\op] (0,\y) circle (\r);
}

\foreach \dx/\dy/\rr in {
 -0.42/4.85/0.06, -0.22/4.60/0.05,  0.28/4.70/0.06,  0.46/4.45/0.05,
 -0.35/4.10/0.06,  0.38/4.10/0.06, -0.18/3.78/0.05,  0.22/3.75/0.05,
 -0.44/3.42/0.06,  0.44/3.42/0.06, -0.25/3.10/0.05,  0.28/3.08/0.05
}{
  \fill[HBARblue!25, opacity=0.23] (\dx,\dy) circle (\rr);
}

\node[HBARblue, font=\bfseries] (Aclab) at (-3.55,4.75) {Atomic cloud};
\draw[-{Latex[length=2.5mm]}, HBARblue, line width=0.9pt] (Aclab.east) -- (-0.20,4.62);

\draw[HBARred, line width=1.2pt, decorate,
      decoration={snake, amplitude=1.2mm, segment length=3.5mm}]
      (0.05,5.55) -- (0.05,6.35);
\draw[HBARred, line width=1.2pt, -{Latex[length=2.8mm]}] (0.05,6.35) -- (0.05,6.65);

\node[HBARred, font=\bfseries, align=center] at (0.35,6.95) {Acceleration\\radiation};

\coordinate (IN) at (2.95,4.60);   
\def\Rin{1.05}                      

\draw[dashed, line width=1.0pt] (IN) circle (\Rin);

\draw[line width=1.2pt] ($(IN)+(-0.62,0.42)$) -- ($(IN)+(0.62,0.42)$);
\draw[line width=1.2pt] ($(IN)+(-0.62,-0.50)$) -- ($(IN)+(0.62,-0.50)$);

\fill[black] ($(IN)+(-0.08,0.42)$) circle (0.075);
\draw[line width=1.1pt] ($(IN)+(-0.08,-0.50)$) circle (0.075);

\draw[HBARblue, line width=1.6pt, -{Latex[length=2.8mm]}]
  ($(IN)+(-0.08,-0.50)$) -- ($(IN)+(-0.08,0.42)$);

\draw[HBARred, line width=1.2pt, decorate,
      decoration={snake, amplitude=1.1mm, segment length=3.0mm}]
      ($(IN)+(0.18,0.55)$) -- ($(IN)+(0.18,1.25)$);
\draw[HBARred, line width=1.2pt, -{Latex[length=2.6mm]}]
      ($(IN)+(0.18,1.25)$) -- ($(IN)+(0.18,1.50)$);

\node[font=\bfseries] at ($(IN)+(0.82,0.42)$) {$|a\rangle$};
\node[font=\bfseries] at ($(IN)+(0.82,-0.50)$) {$|b\rangle$};

\draw[HBARblue, line width=1.0pt] (0.10,4.62) -- ($(IN)+(-\Rin,0.18)$);
\draw[HBARblue, line width=1.0pt] (0.10,4.22) -- ($(IN)+(-\Rin,-0.18)$);

\end{tikzpicture}
\caption{Schematic illustration of horizon-brightened acceleration radiation (HBAR): a dilute atomic cloud falls toward a black hole; a near-horizon mirror suppresses background emission, and atom–field interactions produce acceleration radiation. Inset: effective two-level transition with photon emission \cite{Scully2018}.}
\label{fig:HBAR_schematic}
\end{figure}

In the horizon-brightened acceleration radiation (HBAR) framework, one considers a stream of localized quantum systems—conveniently modeled as two-level atoms—falling toward a black hole and interacting with field modes in the near-horizon region as seen in Fig. \ref{fig:HBAR_schematic} \cite{Scully2018, Camblong2020}. The key physical idea is that near the horizon the kinematics of freely falling matter, together with the strong gravitational redshift, naturally organizes the interaction with the field in a way closely analogous to the Unruh acceleration radiation \cite{Unruh1976, Davies1975}. In this picture, excitation-and-emission processes that would ordinarily remain virtual can be rendered effectively real by the horizon environment, producing radiation with thermal characteristics and an associated entropy flux \cite{Scully2018, Camblong2020}. Such an operational viewpoint is valuable for two reasons: it provides an alternative route to horizon thermodynamics that emphasizes concrete quantum transitions, and it offers a framework that can be exported to geometries beyond the simplest Schwarzschild case, thereby testing how universal the thermodynamic interpretation remains \cite{Camblong2020}. Recent progress has substantially broadened this quantum-optical viewpoint on horizon physics. 
A unified perspective emphasizing the interplay between acceleration radiation, black holes, and quantum optics has been developed in a series of works that connect near-horizon universality to conformal quantum mechanics and related open-system formulations \cite{Ordonez:2025sqp,Azizi:2021yto,Chakraborty:2022qdr}. 
In particular, conformal quantum mechanics has emerged as a useful organizing principle not only in black-hole settings but also in causal-diamond thermality and instability, providing a complementary route to effective horizon dynamics \cite{Camblong:2024jpq}. 
On the operational side, the possibility of enhancing acceleration radiation through cavity QED mechanisms highlights that detector--field engineering can strongly influence emission and entropy production \cite{Scully:2003zz}. Motivated by these developments, HBAR-inspired analyses have been extended to increasingly realistic and diverse gravitational environments, including derivative-coupled atoms and modified-gravity black holes \cite{Pantig:2025okn}, quantum-corrected charged geometries and associated entropy diagnostics \cite{Jana:2024fhx,Jana:2025hfl,Sen:2022tru}, and braneworld backgrounds \cite{Das:2023rwg}. 
Related studies have explored how additional sectors such as dark-energy or dark-matter phenomenology can imprint on acceleration-radiation observables \cite{Bukhari:2022wyx,Bukhari:2023yuy}, and how Lorentz-violating effects might be constrained through acceleration-radiation signatures \cite{Tang:2025eew}. Within this broader program, the HBAR entropy flux has also been analyzed in quantum-gravity motivated deformations of near-horizon physics, including a GUP-corrected Schwarzschild geometry, where the robustness of thermality and the role of the equivalence principle can be assessed in a controlled setting \cite{Ovgun:2025isv}. Finally, the ongoing effort to connect these ideas to empirical access has stimulated both conceptual discussions and concrete experimental strategies, including proposals involving accelerated charges/electrons \cite{Gregori:2023tun} and broader quantum-information aspects of Unruh/Hawking-like processes \cite{Scully:2022pov,Scully:2025rro}. 
In parallel, complementary spectroscopic probes of strong-field geometry from orbit-based quantization ideas to shadow-based diagnostics continue to enrich the observational context in which horizon-related quantum effects may ultimately be assessed \cite{Pantig:2025zhn}.


The purpose of the present work is to synthesize these lines of inquiry by applying the HBAR formalism to a broad class of generic regular black holes sourced by nonlinear electrodynamics, while simultaneously developing the corresponding optical diagnostics in the same spacetimes. We focus on static, spherically symmetric, asymptotically well-behaved regular geometries that interpolate between familiar limiting cases and capture a range of nonlinear electrodynamic effects through a small set of parameters. This “generic” description is not merely aesthetic: it allows us to isolate which aspects of the HBAR construction depend only on near-horizon universality (such as the local Rindler structure) and which aspects are sensitive to the global geometry (such as the mapping between horizon data, geodesic structure, and asymptotic observers). In nonlinear electrodynamics, an additional subtlety arises because light propagation can be governed by an effective optical geometry rather than the original spacetime metric; accordingly, careful treatment of null trajectories is essential when connecting to shadow and lensing observables \cite{Novello2000, Obukhov2002}.

We adopt the following conventions throughout. We work in four spacetime dimensions and use the metric signature $(-,+,+,+)$. We employ relativistic units where $G = c = 1$, while keeping the reduced Planck constant explicit whenever quantum transition rates, thermal factors, or entropy fluxes are discussed. Spacetime indices are raised and lowered with the spacetime metric, and we use standard curvature conventions common in the general-relativity literature. Geometrically, we use an areal radial coordinate so that spheres of constant radius have the usual area, and we restrict attention to the exterior region relevant for asymptotic observations as well as the near-horizon region relevant for the HBAR construction. On the matter side, the nonlinear electrodynamics sector is characterized by a Lagrangian depending on the usual electromagnetic invariant(s), and we treat the infalling microscopic systems as idealized two-level atoms coupled to a field mode in a way that permits a controlled quantum-statistical description of radiation and entropy production.

This paper is organized as follows: In Sec. \ref{sec2} we present the generic regular-black-hole geometries arising in general relativity coupled to nonlinear electrodynamics and specify the assumptions that define the class under study. In Sec. \ref{sec3} and \ref{sec4} we derive the relevant geodesic structure and the associated optical diagnostics, including the quantities needed to characterize the shadow radius and the weak-field deflection angle as seen by distant observers. In Sec. \ref{sec5} we develop the near-horizon reduction that casts the dynamics into the form appropriate for a conformal quantum mechanics description, preparing the ground for the HBAR analysis. In Sec. \ref{sec6} we apply the HBAR formalism to atoms falling into these generic regular black holes and compute the corresponding radiation characteristics in a way suited to entropy accounting. In Sec. \ref{sec7} we formulate the HBAR entropy flux and analyze its connection to the horizon area law within the present class of spacetimes. We also explored the HBAR entropy and the area law in Sec. \ref{sec8}, and the Wien's displacement in Sec. \ref{sec9}. We end with a brief discussion of implications and limitations of the framework in the context of regular black holes supported by nonlinear electrodynamics in Sec. \ref{sec10}.

\section{THE SPACE-TIME METRIC OF THE
GENERIC REGULAR BLACK HOLES} \label{sec2}
\begin{figure*}[t]
\includegraphics[width=0.45\textwidth]{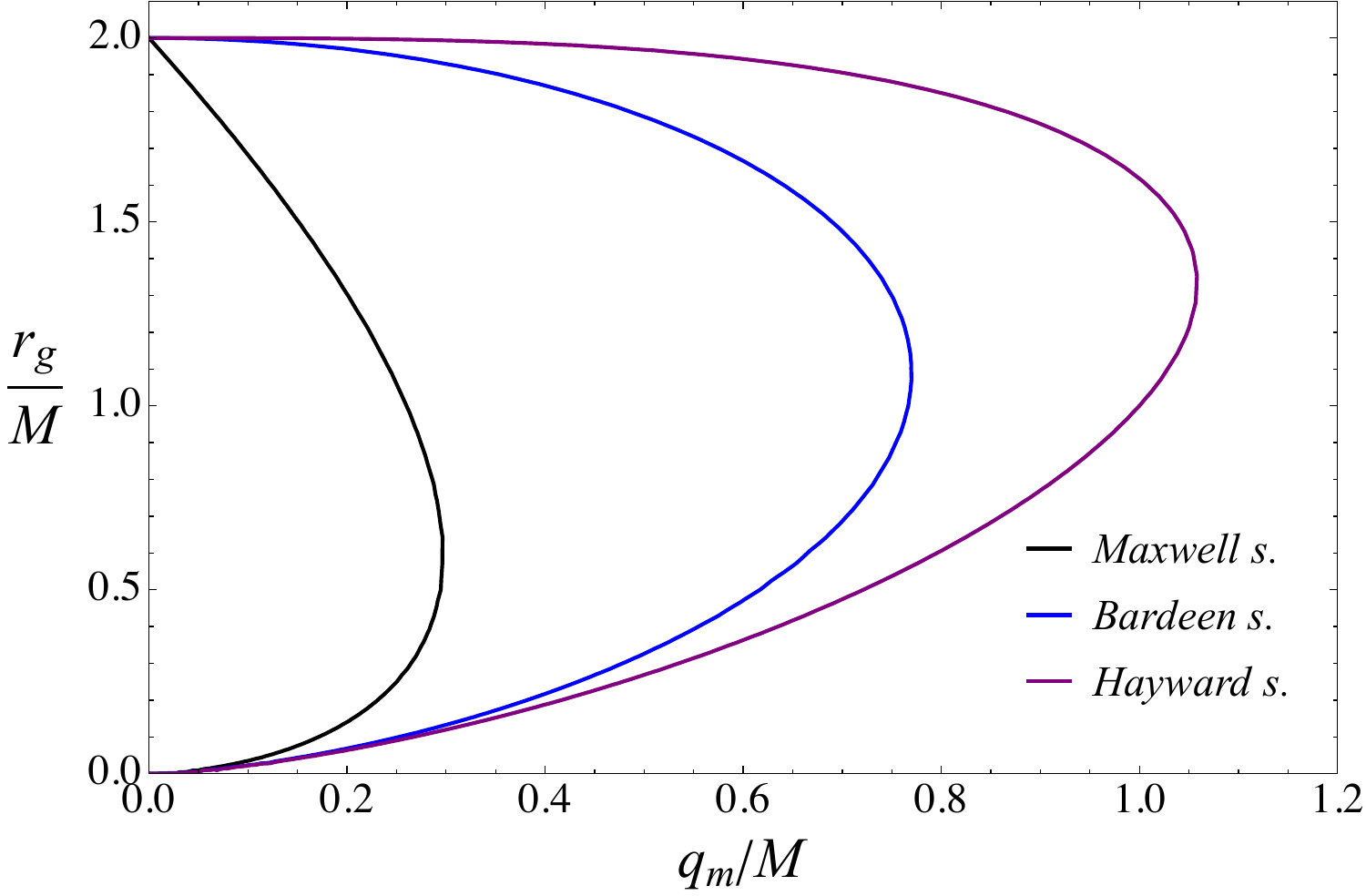}
\includegraphics[width=0.45\textwidth]{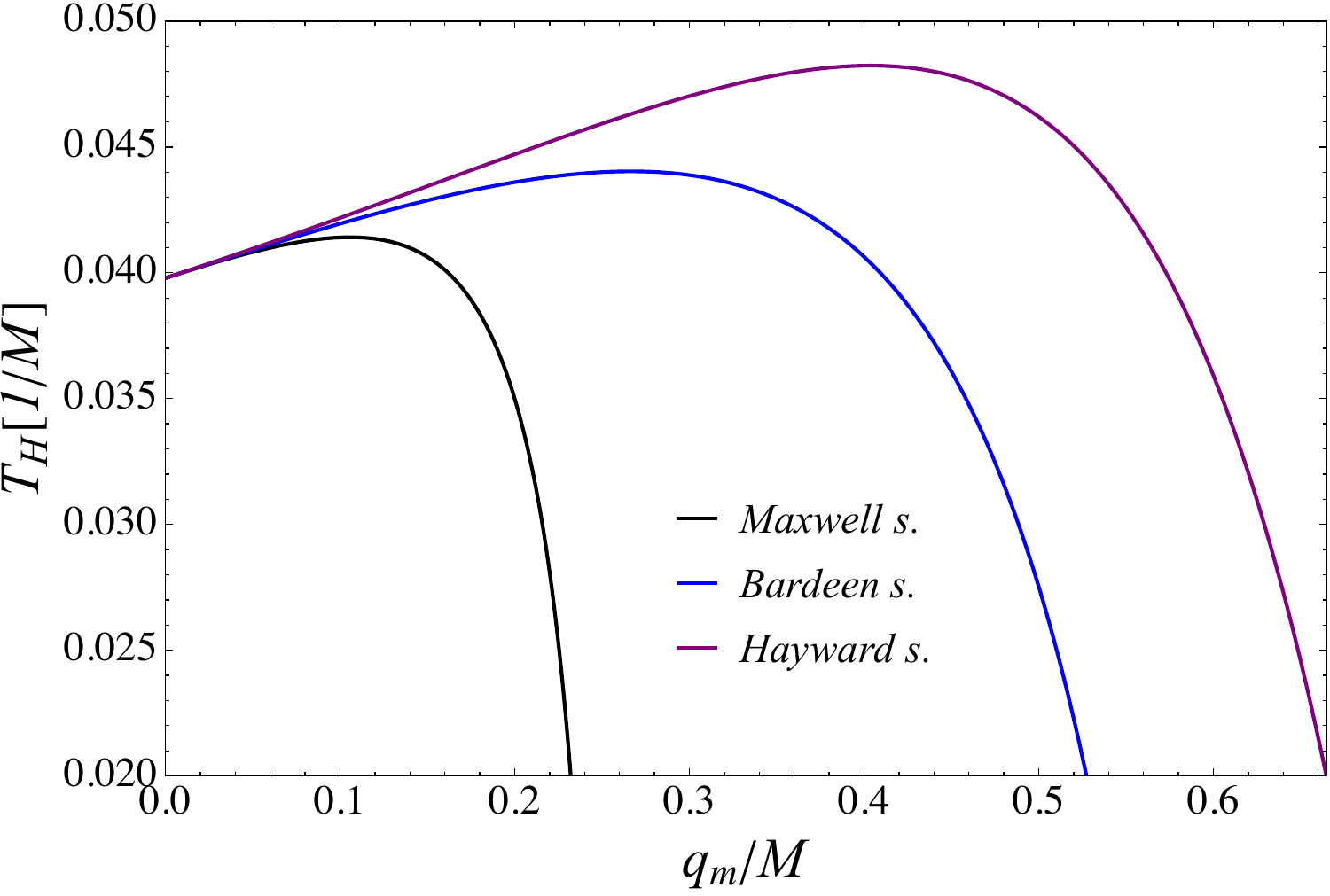}
\caption{Dependence of the horizon radius $r_g$ (left panel) and Hawking temperature $T_H$ (right panel) on magnetic charge $q_m$ for different solutions.}
    \label{fig:rg and T}
\end{figure*}

The Lagrangian density describing $\mathcal{L}$ the nonlinear electrodynamics (NED) alters the action as (\cite{Toshmatov2019}):
\begin{eqnarray}\label{eq.action}
    S=\frac{1}{16\pi}\int d^4x\sqrt{-g}\left[R-\mathcal{L}(F)\right]\,,
\end{eqnarray}
where $R$ is the Ricci scalar and $g$ is the determinant of the space-time metric. Faraday scalar $F=F_{\mu\nu}F^{\mu\nu}$, in which  $F_{\mu\nu}=\partial_\mu A_\nu-\partial_\nu A_\mu$ is the electromagnetic tensor with the electromagnetic four potential $A^\mu$, allows to find a regular black hole solutions without spacetime singularities \cite{Toshmatov2019}:
\begin{subequations}\label{eq.the metric}
\begin{align}
    &ds^2=-f(r)dt^2+\frac{dr^2}{f(r)}+r^2d\Omega^2\,,\\
    &f(r)=1-\frac{2Mr^{\mu-1}}{\left(r^{\nu}+q_m^\nu\right)^{\frac{\mu}{\nu}}}
\end{align}
\end{subequations}
 in which $d\Omega^2=\sin^2{\theta}d\phi^2+d\theta^2$ is the usual Schwarzschild-like coordinates. Fan \& Wang have shown that this Black hole spacetime (BH) is regular when condition $\mu\geq3$ is fulfilled (\cite{2016PhRvD..94l4027F}). Also, the line-element in Eq.(\ref{eq.the metric}) can describe several well-known BH solutions such as, Maxwellian solution $\nu=1$, Bardeen-like solution $\nu=2$ and Hayward-like solution $\nu=3$. In this study, we consider these three solutions with minimal parameter $\mu=3$. To find horizons of the space-time is important task for our further calculation which is for our case $f(r_g)=0$. Then in Fig.(\ref{fig:rg and T}) we depict how horizon radius depends on the magnetic charge $q_m$ for different solutions  with $\mu=3$. One can see from Fig.(\ref{fig:rg and T}) that there are two horizons inner and outer which coincides for extremal values $q^*_m$. This extreme value $q^*_m$ can be obtained by solving Eqs. $f(r)=\frac{df(r)}{dr}=0$, simultaneously which are given in Table (\ref{Table 1}).

It is clear from fig.(\ref{fig:rg and T}) that BHs exist only when $q_m\leq q_m^*$ and Hayward-like BH solutions have large event horizon than Maxwell BHs and Bardeen BHs.

\begin{table}[ht!]
    \centering
    \begin{tabular}{|c|c|c|}
     \hline
       $\nu$ & $q_m^*[M]$ & $r_g^*[M]$  \\
    \hline
      1   & 0.296296   &0.592593 \\
      2   & 0.7698   &1.08866 \\ 
      3   & 1.05827   &1.33333  \\ 
   \hline
    \end{tabular}
    \caption{Extremal values $q_m^*$ and corresponding extremal values $r_g^*$ for different solutions with $\mu=3$}
    \label{Table 1}
\end{table}

Then surface gravity $\kappa$ and Hawking temperature $T_H$ of the generic regular BH can be expressed as (\cite{Ovgun:2025ehi}) :
\begin{subequations}\label{eq.k and T}
    \begin{align}
&\kappa=\frac{f'(r_g)}{2}=\frac{\left(r_g^\nu-2q_m^\nu\right)}{2r_g\left(r_g^\nu+q^\nu\right)}\,,\\
&T_H=\frac{\kappa}{2\pi}=\frac{\left(r_g^\nu-2q_m^\nu\right)}{4\pi r_g\left(r_g^\nu+q_m^\nu\right)}\,,
    \end{align}
\end{subequations}

here we have used expression (with $\mu=3$) $M=\frac{\left(r_g^\nu+q_m^\nu\right)^{\frac{3}{\nu}}}{2r_g^2}$ .

Following, we have depicted dependence of the Hawking temperature $T_H$ on the magnetic charge $q_m$ for different solutions in Fig.(\ref{fig:rg and T}). The general nature of the Hawking temperature $T_H$ is similar for all solutions: initially growing till some values of the $q_m$ then decreasing monotonically, also it is clear from Fig.(\ref{fig:rg and T}) that the highest value of the $T_H$ corresponds to the Hayward-like BH solution.

\section{SHADOW ANALYSIS} \label{sec3}
We analyze the gravitational shadow cast by the static, spherically symmetric spacetime \cite{Pantig:2025deu,Kobialko:2024zhc}
\begin{equation}
ds^2=-\alpha(r)\,dt^2+\gamma(r)\,dr^2+\beta(r)\left(d\theta^2+\sin^2\theta\,d\phi^2\right),
\label{3.1}
\end{equation}
specialized to the line element
\begin{eqnarray}
\alpha(r)=f(r),\qquad \gamma(r)=\frac{1}{f(r)},\qquad \beta(r)=r^2,
\\ \notag
f(r)=1-\frac{2Mr^{\mu-1}}{\left(r^{\nu}+q_m^\nu\right)^{\mu/\nu}}.
\label{3.2}
\end{eqnarray}
Here $M$ is the mass parameter, $q_m$ is the magnetic charge parameter, and $(\mu,\nu)$ are dimensionless model parameters. Throughout this section we work in geometric units $G=c=1$. We compute the shadow boundary for massless probes propagating along null geodesics of the spacetime metric (geometric-optics approximation), following the tetrad-based formalism for static spherically symmetric geometries.

Because of spherical symmetry we restrict to the equatorial plane $\theta=\pi/2$ without loss of generality. For a geodesic $x^\mu(\lambda)$ parametrized by an affine parameter $\lambda$, the Killing vectors $\partial_t$ and $\partial_\phi$ yield the conserved energy $E$ and angular momentum $L$,
\begin{equation}
E=\alpha(r)\,\dot t,\qquad L=\beta(r)\,\dot\phi,
\label{3.3}
\end{equation}
where an overdot denotes $d/d\lambda$. The normalization condition $g_{\mu\nu}\dot x^\mu\dot x^\nu=-m^2$ gives the radial equation \cite{Pantig:2025deu}
\begin{equation}
\gamma(r)\,\dot r^{\,2}= \frac{E^2}{\alpha(r)}-\frac{L^2}{\beta(r)}-m^2.
\label{3.4}
\end{equation}
It is convenient to introduce the dimensionless parameters (in the notation adapted to the shadow formalism)
\begin{equation}
\epsilon\equiv \frac{m^2}{E^2},\qquad l\equiv \frac{L^2}{E^2}.
\label{3.5}
\end{equation}
Dividing Eq. \eqref{3.4} by $E^2$ and evaluating at a radial turning point (where $\dot r=0$) yields the algebraic relation between $(\epsilon,l)$ and the turning-point radius,
\begin{equation}
\epsilon=\frac{1}{\alpha(r)}-\frac{l}{\beta(r)}.
\label{3.6}
\end{equation}
The shadow boundary corresponds to critical scattering trajectories whose turning point becomes degenerate (the effective potential has a double root). Differentiating Eq. \eqref{3.6} with respect to $r$ at fixed $(\epsilon,l)$ and imposing the degeneracy condition leads to \cite{Pantig:2025deu}
\begin{equation}
-\frac{\alpha_{,r}(r)}{\alpha(r)^2}+\frac{\beta_{,r}(r)}{\beta(r)^2}\,l=0,
\label{3.7}
\end{equation}
which can be solved for $l$ at the critical radius. The observed angular radius of the shadow can be extracted from a local orthonormal tetrad of a static observer at a finite observation point $\bar r$; for an equatorial observer one finds the invariant relation
\begin{equation}
\sin^2\Theta=\frac{\bar\alpha}{\alpha(r)}\cdot\frac{\beta(r)}{\bar\beta}\cdot
\frac{1-\alpha(r)\epsilon}{1-\bar\alpha\,\epsilon},
\qquad \bar\alpha\equiv \alpha(\bar r),\quad \bar\beta\equiv \beta(\bar r),
\label{3.8}
\end{equation}
where $\Theta$ is the angle on the observer's celestial sphere measured from the radial direction. For asymptotically flat spacetimes and an asymptotic static observer $\bar r\to\infty$, we have $\bar\alpha\to 1$ and $\bar\beta\sim \bar r^2$, and it is natural to define the finite shadow radius by
\begin{equation}
R\equiv \lim_{\bar r\to\infty}\bar r\,\Theta.
\label{3.9}
\end{equation}
Combining Eqs. \eqref{3.6}-\eqref{3.9} gives a compact expression for $R^2$ in terms of $\alpha(r)$ and $\beta(r)$ evaluated at the critical (degenerate turning-point) radius. In particular, for null geodesics ($m=0$ so that $\epsilon=0$) one obtains \cite{Pantig:2025deu}
\begin{equation}
R^2=\frac{\beta(r_{\rm ph})}{\alpha(r_{\rm ph})},
\qquad
\left[\beta\,\alpha_{,r}-\alpha\,\beta_{,r}\right]_{r=r_{\rm ph}}=0,
\label{3.10}
\end{equation}
where $r_{\rm ph}$ is the radius of the unstable circular null orbit (photon sphere) that forms the shadow boundary for an asymptotic observer.

For our metric Eq. \eqref{3.2} we have $\alpha(r)=f(r)$ and $\beta(r)=r^2$, hence Eq. \eqref{3.10} becomes the standard photon-sphere condition
\begin{equation}
\left[2f(r)-r f'(r)\right]_{r=r_{\rm ph}}=0,
\label{3.11}
\end{equation}
and the shadow radius reduces to
\begin{equation}
R_{\rm sh}^2=\frac{r_{\rm ph}^2}{f(r_{\rm ph})}.
\label{3.12}
\end{equation}
To apply Eq. \eqref{3.11} explicitly, we first compute the derivative of $f(r)$:
\begin{equation}
f'(r)=
2M\,r^{\mu-2}\left(r^{\nu}+q_m^\nu\right)^{-\mu/\nu-1}
\left[r^{\nu}-(\mu-1)q_m^\nu\right].
\label{3.13}
\end{equation}
Substituting Eqs. \eqref{3.2} and \eqref{3.13} into Eq. \eqref{3.11} and simplifying yields an implicit equation for the photon-sphere radius,
\begin{equation}
\left(r_{\rm ph}^{\nu}+q_m^\nu\right)^{\mu/\nu+1}
=
M\,r_{\rm ph}^{\mu-1}\left[3r_{\rm ph}^{\nu}+(3-\mu)q_m^\nu\right].
\label{3.14}
\end{equation}
The physically relevant solution is the largest positive root of Eq. \eqref{3.14} lying outside the event horizon (when a horizon exists), corresponding to the unstable outer photon sphere.

It is often useful to eliminate $M$ from $f(r_{\rm ph})$ using Eq. \eqref{3.14}. Solving Eq. \eqref{3.14} for the combination $Mr_{\rm ph}^{\mu-1}\left(r_{\rm ph}^{\nu}+q_m^\nu\right)^{-\mu/\nu}$ and substituting into $f(r_{\rm ph})$ gives
\begin{equation}
f(r_{\rm ph})=
\frac{r_{\rm ph}^{\nu}+(1-\mu)q_m^\nu}{3r_{\rm ph}^{\nu}+(3-\mu)q_m^\nu},
\label{3.15}
\end{equation}
and therefore Eq. \eqref{3.12} can be written equivalently as
\begin{equation}
R_{\rm sh}^2
=
r_{\rm ph}^2\,
\frac{3r_{\rm ph}^{\nu}+(3-\mu)q_m^\nu}{r_{\rm ph}^{\nu}+(1-\mu)q_m^\nu}.
\label{3.16}
\end{equation}
For the regular subclass $\mu=3$ frequently used in nonlinear-electrodynamics black holes, Eq. \eqref{3.14} simplifies to
\begin{equation}
\left(r_{\rm ph}^{\nu}+q_m^\nu\right)^{(\nu+3)/\nu}=3M\,r_{\rm ph}^{\nu+2},
\label{3.17}
\end{equation}
while Eqs. \eqref{3.15} and \eqref{3.16} reduce to
\begin{equation}
f(r_{\rm ph})=\frac{r_{\rm ph}^{\nu}-2q_m^\nu}{3r_{\rm ph}^{\nu}},
\qquad
R_{\rm sh}^2=\frac{3\,r_{\rm ph}^{\nu+2}}{r_{\rm ph}^{\nu}-2q_m^\nu}.
\label{3.18}
\end{equation}
As a consistency check, in the Schwarzschild limit $q_m\to 0$ Eq. \eqref{3.17} gives $r_{\rm ph}=3M$, and Eq. \eqref{3.12} yields the standard shadow radius $R_{\rm sh}=3\sqrt{3}\,M$.

We emphasize that Eq. \eqref{3.14} is, in general, a transcendental algebraic condition for $r_{\rm ph}$ once $(\mu,\nu)$ are kept arbitrary. While closed-form expressions in radicals may exist for special integer choices (e.g.\ cases reducing to cubic or quartic equations), the resulting formulas are typically unwieldy and do not provide additional physical transparency. For this reason it is standard to treat $r_{\rm ph}$ as the largest positive root of Eq. \eqref{3.14} and then substitute into Eqs. \eqref{3.12} or \eqref{3.16}. Nevertheless, one can render the parameter dependence explicit by scaling out the mass and, when desired, develop controlled analytic approximations.

To make the mass scaling manifest, we introduce the dimensionless variables
\begin{equation}
x\equiv \frac{r}{M},\qquad Q\equiv \frac{q_m}{M}.
\label{3.19}
\end{equation}
In terms of $(x,Q)$, the photon-sphere condition Eq. \eqref{3.14} becomes an equation independent of $M$,
\begin{equation}
\left(x_{\rm ph}^{\nu}+Q^\nu\right)^{\mu/\nu+1}
=
x_{\rm ph}^{\mu-1}\left[3x_{\rm ph}^{\nu}+(3-\mu)Q^\nu\right],
\label{3.20}
\end{equation}
whose physically relevant solution is the largest real root $x_{\rm ph}>0$ outside the horizon (when present). The shadow radius then factorizes as
\begin{eqnarray}
R_{\rm sh}=M\,\mathcal{R}_{\rm sh}(\mu,\nu;Q),
\qquad 
\mathcal{R}_{\rm sh}^2(\mu,\nu;Q)=\frac{x_{\rm ph}^2}{\tilde f(x_{\rm ph})},
\\
\tilde f(x)\equiv 1-\frac{2x^{\mu-1}}{\left(x^{\nu}+Q^\nu\right)^{\mu/\nu}},
\label{3.21}
\end{eqnarray}
which is equivalent to Eq. \eqref{3.16} upon using Eq. \eqref{3.20}.

In the regime $Q^\nu\ll 1$, one may obtain an explicit analytic approximation by expanding about the Schwarzschild photon sphere $x_{\rm ph}=3$. Solving Eq. \eqref{3.20} perturbatively yields
\begin{equation}
x_{\rm ph}
=
3-\frac{\mu(\nu+3)}{\nu\,3^{\nu}}\,Q^\nu+\mathcal{O}\!\left(Q^{2\nu}\right),
\label{3.22}
\end{equation}
so that, in terms of the original parameters,
\begin{equation}
r_{\rm ph}
=
3M\left[
1-\frac{\mu(\nu+3)}{\nu\,3^{\nu+1}}\left(\frac{q_m}{M}\right)^\nu
+\mathcal{O}\!\left(\left(\frac{q_m}{M}\right)^{2\nu}\right)
\right].
\label{3.23}
\end{equation}
Substituting Eq. \eqref{3.22} into Eq. \eqref{3.21} and expanding consistently gives the shadow radius as an explicit function of $M$ and the nonlinear-electrodynamics parameters,
\begin{equation}
R_{\rm sh}
=
3\sqrt{3}\,M\left[
1-\frac{\mu(\nu+3)}{\nu\,3^{\nu+1}}\left(\frac{q_m}{M}\right)^\nu
+\mathcal{O}\!\left(\left(\frac{q_m}{M}\right)^{2\nu}\right)
\right],
\label{3.24}
\end{equation}
or equivalently, at the level of $R_{\rm sh}^2$,
\begin{equation}
R_{\rm sh}^2
=
27\,M^2\left[
1-\frac{2\mu(\nu+3)}{\nu\,3^{\nu+1}}\left(\frac{q_m}{M}\right)^\nu
+\mathcal{O}\!\left(\left(\frac{q_m}{M}\right)^{2\nu}\right)
\right].
\label{3.25}
\end{equation}
For the regular subclass $\mu=3$, Eqs. \eqref{3.23} and \eqref{3.24} simplify to
\begin{eqnarray}\label{eq.shadow}
r_{\rm ph}
=
3M\left[
1-\frac{\nu+3}{\nu\,3^{\nu}}\left(\frac{q_m}{M}\right)^\nu
+\mathcal{O}\!\left(\left(\frac{q_m}{M}\right)^{2\nu}\right)
\right],
\\ \notag
R_{\rm sh}
=
3\sqrt{3}\,M\left[
1-\frac{\nu+3}{\nu\,3^{\nu}}\left(\frac{q_m}{M}\right)^\nu
+\mathcal{O}\!\left(\left(\frac{q_m}{M}\right)^{2\nu}\right)
\right].
\label{3.26}
\end{eqnarray}
These expressions make explicit how the deviation from the Schwarzschild shadow size is controlled, at leading order, by the dimensionless ratio $(q_m/M)^\nu$ and by the model parameters $(\mu,\nu)$. For general values of $(\mu,\nu;Q)$ beyond the perturbative regime, one determines $x_{\rm ph}$ by solving Eq. \eqref{3.20} and then evaluates $R_{\rm sh}$ through Eq. \eqref{3.21} (or, equivalently, through Eq. \eqref{3.16}).

\begin{figure*}[t]
\includegraphics[width=0.45\textwidth]{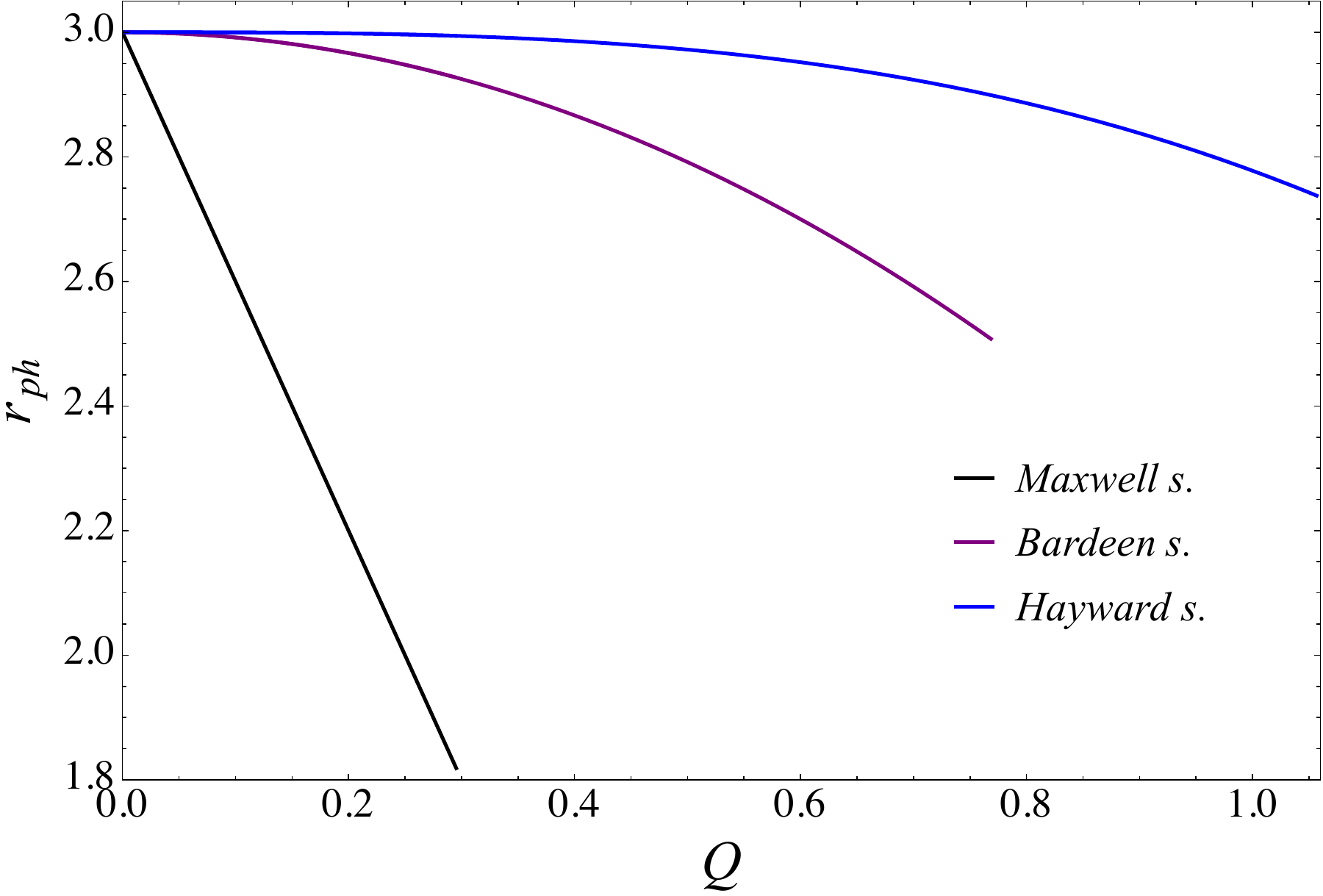}
\caption{Dependence of the photon radius $r_{ph}$  on dimensionless magnetic charge $Q=\frac{q_m}{M}$ for different solutions.}
    \label{fig:rph}
\end{figure*}

\begin{figure*}[t]
\includegraphics[width=0.45\textwidth]{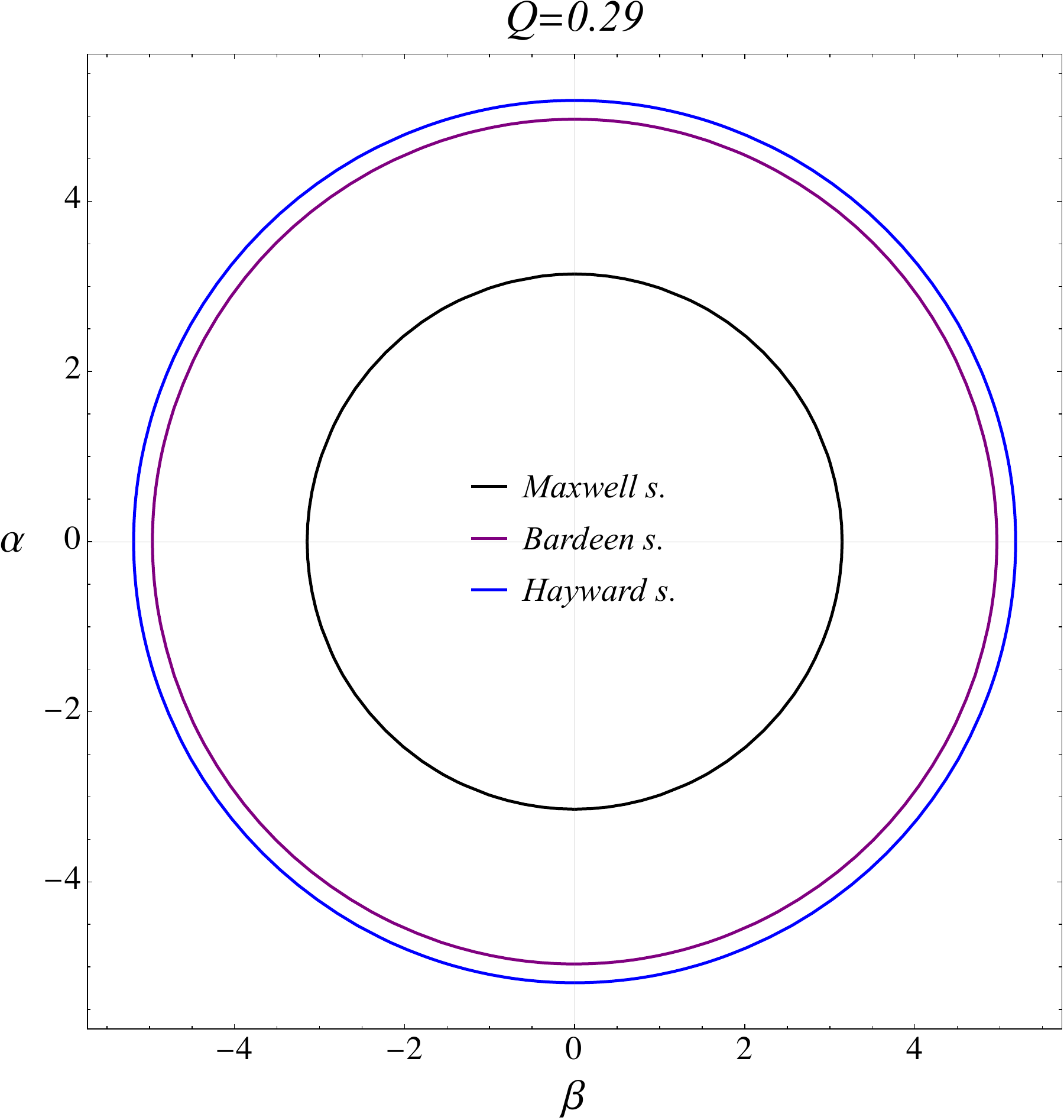}
\includegraphics[width=0.45\textwidth]{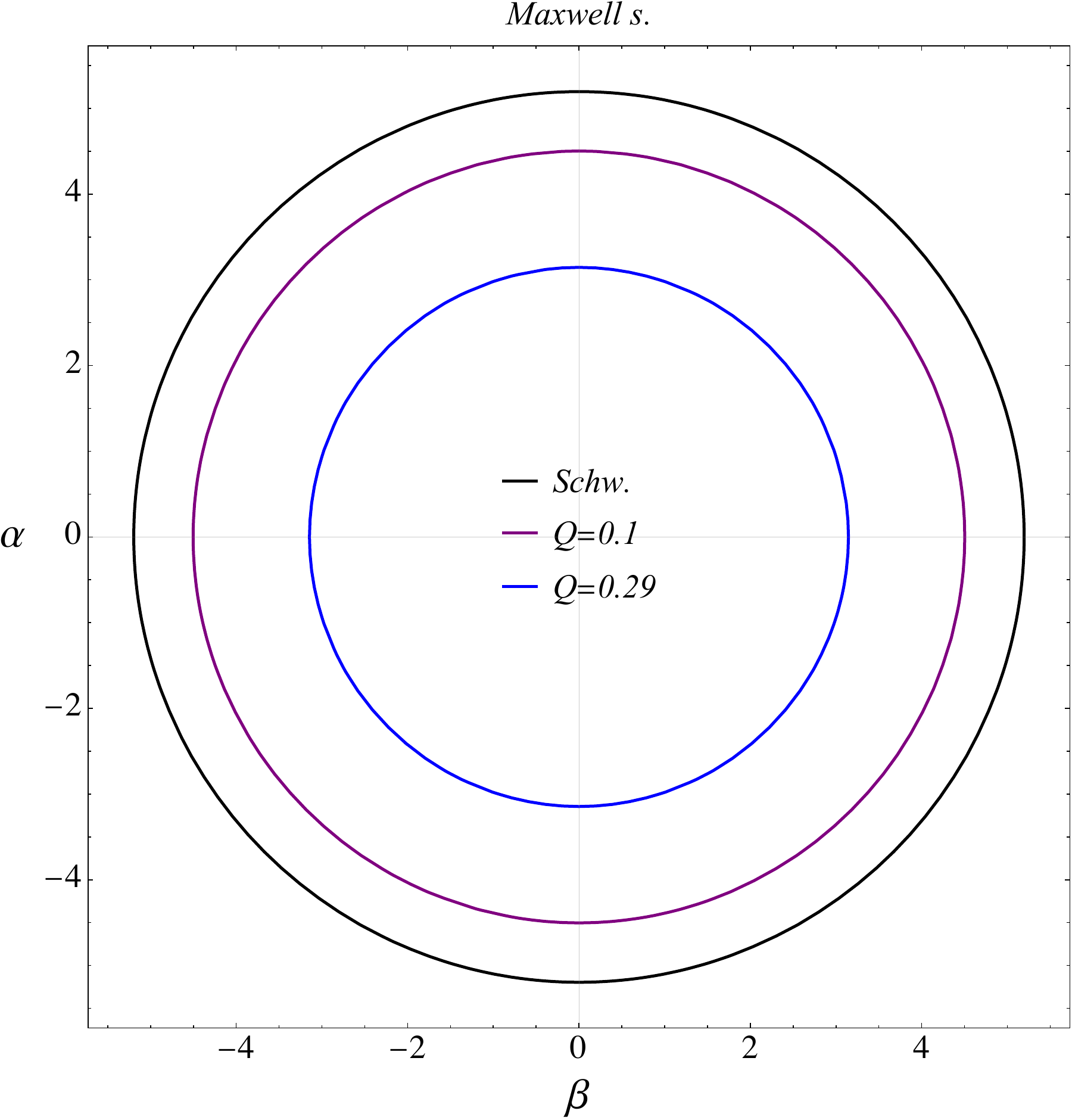}
\includegraphics[width=0.45\textwidth]{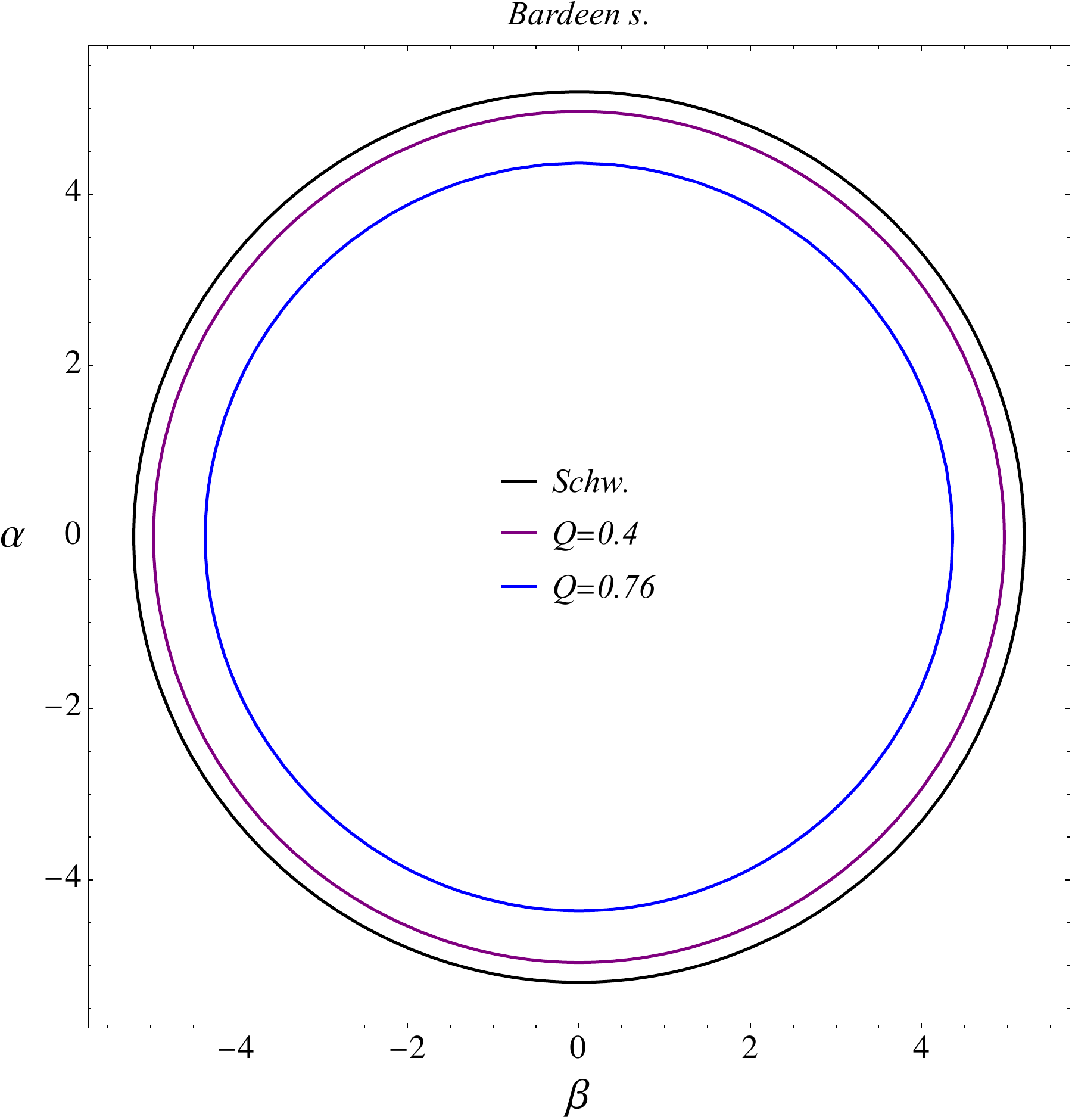}
\includegraphics[width=0.45\textwidth]{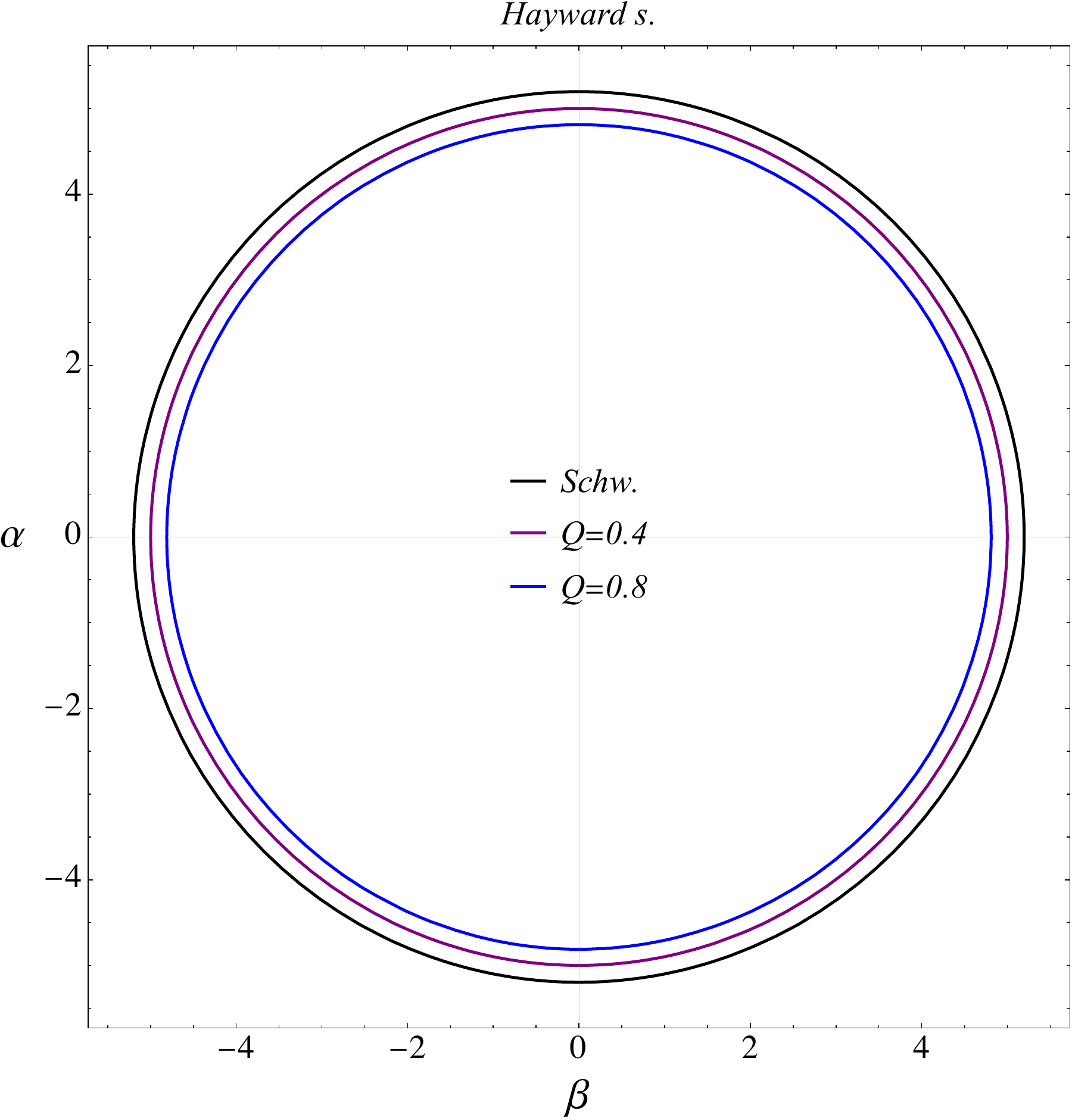}
\caption{The shadow image for different cases}
    \label{fig:Rsh}
\end{figure*}

Then we have plotted dependence of the photon radius $r_{ph}$ on the dimensionless magnetic charge $Q=\frac{q_m}{M}$ for different solutions in Fig.(\ref{fig:rph}). One can conclude from Fig.(\ref{fig:rph}) that increasing magnetic charge $Q$ causes decreasing the value of the photon radius $r_{ph}$ for all solutions and it is clear that the effect of $Q$ is the strongest for Maxwell solution. Additionally, the image of the shadow of the generic regular BHs is shown in Fig.(\ref{fig:Rsh}). Increasing the magnetic charge $Q$ leads shrinking the shadow of the generic regular BHs.

\section{PARAMETER ESTIMATION FOR GENERIC REGULAR BLACK HOLES VIA MCMC ANALYSIS} \label{sec4}
Now employing Python library emcee package and using observed data for $Sgr A^*$ from Event Horizon Telescope (EHT) (\cite{EventHorizonTelescope:2022wkp}),  GRAVITY collaboration (\cite{GRAVITY:2021xju}) and observed  data for $M 87^*$ from  EHT collaboration (\cite{EventHorizonTelescope:2025vum}) we will employ Markov Chain Monte Carlo (MCMC) analysis to constrain NED parameters $\nu$, $Q=\frac{q}{M}$ and BH mass $M$.

Posterior distribution can be expressed as (\cite{Abdulkhamidov:2024lvp}):
\begin{eqnarray}\label{eq.posterior distribution}
P(\Theta|\mathcal{D},\mathcal{M})=\frac{\mathcal{L}(\Theta|\mathcal{D},\mathcal{M})\pi(\Theta|\mathcal{M})}{P(\mathcal{D}|\mathcal{M})},
\end{eqnarray}
where $\mathcal{L}\left(\Theta|\mathcal{D},\mathcal{M}\right)$ is the likelihood of the parameters $\Theta$ given data $\mathcal{D}$ and model $\mathcal{M}$. This is combined with the prior $\pi\left(\Theta|\mathcal{M}\right)$ 
 and normalized by the marginal likelihood $P\left(\mathcal{D}|\mathcal{M}\right)$ to compute the posterior distribution. The priors are set to be Gaussian priors, with boundary condition:
 \begin{eqnarray}\label{eq.Gaussian}
\pi(\Theta_i)\sim\exp{\left[\left(\frac{\Theta_i-\Theta_{0,i}}{\sigma_i}\right)^2\right]},
\end{eqnarray}
where $\sigma_i$ is the standard deviation. Also, we give numerical values for the priors  of the parameters of the generic regular BHs used in our analysis in Table.(\ref{Table 2}). 

\begin{table}[h!]
\centering
\resizebox{.5\textwidth}{!}{
\begin{tabular}{|c|cc|cc|cc|cc|cc|}
\hline
& \multicolumn{2}{c|}{$Sgr A^*$ from EHT} & \multicolumn{2}{c|}{$Sgr A^*$ from Gravity} & \multicolumn{2}{c|}{$M 87^*$ from EHT}  \\
 & $\mu$ & $\sigma$ & $\mu$ & $\sigma$ & $\mu$ & $\sigma$ \\
\hline
$M \ (\times 10^6M_\odot)$ & 4.25 & 0.6 & 4.1 & 0.6 & 6300 & 250  \\
$\nu$ & 1.13 & 0.1 & 1.044 & 0.2 & 1.621 & 0.31  \\
$Q$ & 0.255 & 0.05 & 0.326 & 0.05 & 0.004 & 0.0007 \\
\hline
\end{tabular}
}
\caption{ Gaussian prior on   generic regular BHs  from angular shadow size constraints.}
\label{Table 2}
\end{table}

Then the likelihood function can be expressed as:
\begin{eqnarray}\label{eq.Lik.}
    \mathcal{L}=-\frac{1}{2}\Sigma_{i}\left(\frac{\theta_{obs.}^i-\theta_{th.}^i}{\sigma_i}\right)^2\,,
\end{eqnarray}
where $\theta_{obs.}^i$ is the observed values of the angular shadow size of the BHs and $\theta_{obs.}^i$ is the theoretical values of the angular shadow size:
\begin{eqnarray}\label{eq.ang.shad.}
    \theta_{obs.}^i=\frac{2R_{sh}}{D}\times\frac{G\alpha}{c^2}\,,
\end{eqnarray}
where $R_{sh}$ is the shadow radius (\ref{eq.shadow}), $D$ is the distance to the black hole, $\alpha=\frac{180\times3600}{\pi}\times10^{6}$ is the converting parameter of the unity from radian into microarcsecond and we have restored numerical values of the gravitational constant $G$ and speed of the light $c$ in Eq.(\ref{eq.ang.shad.}) to coincide our theoretical results with observed data.
\begin{figure*}[t]
\includegraphics[width=0.45\textwidth]{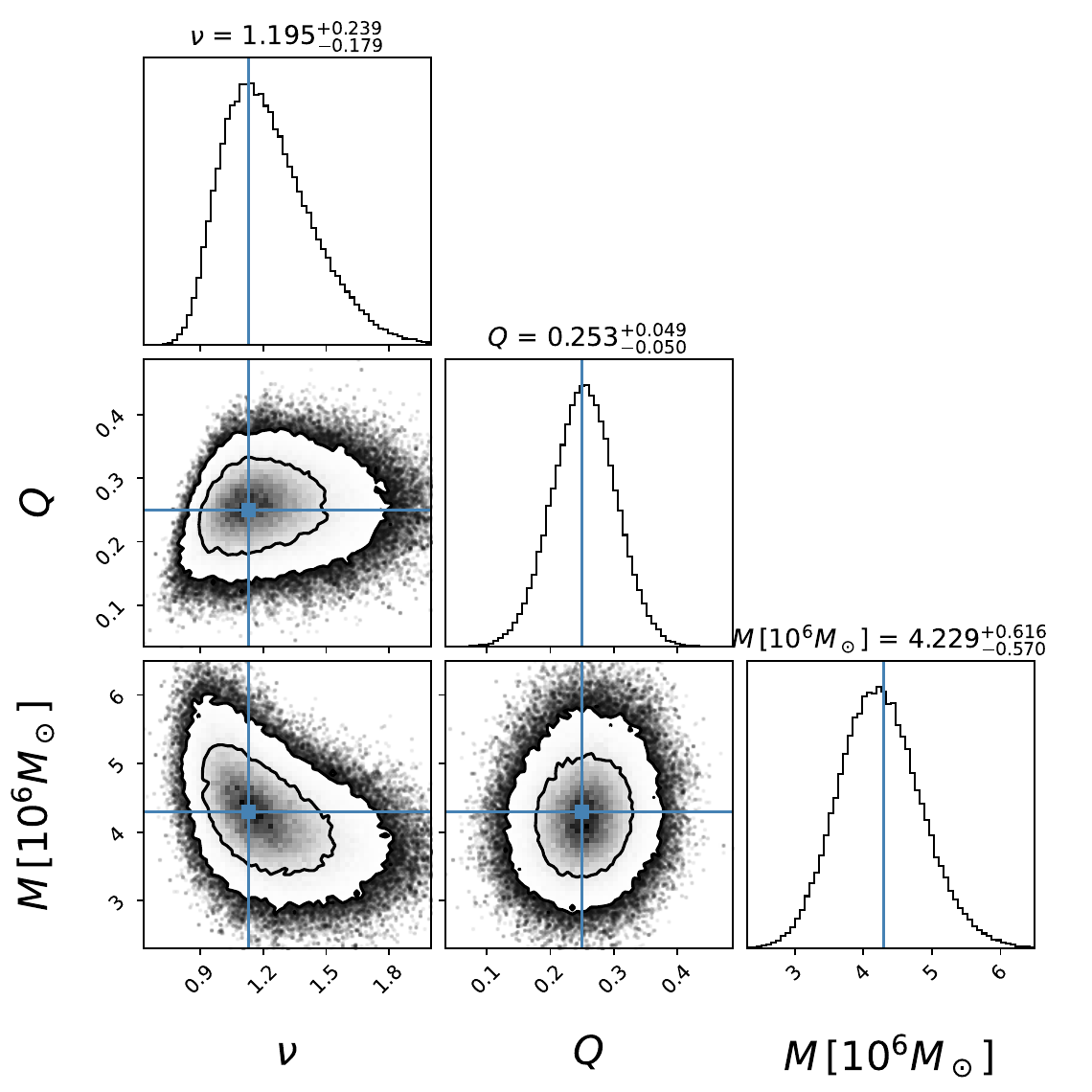}
\includegraphics[width=0.45\textwidth]{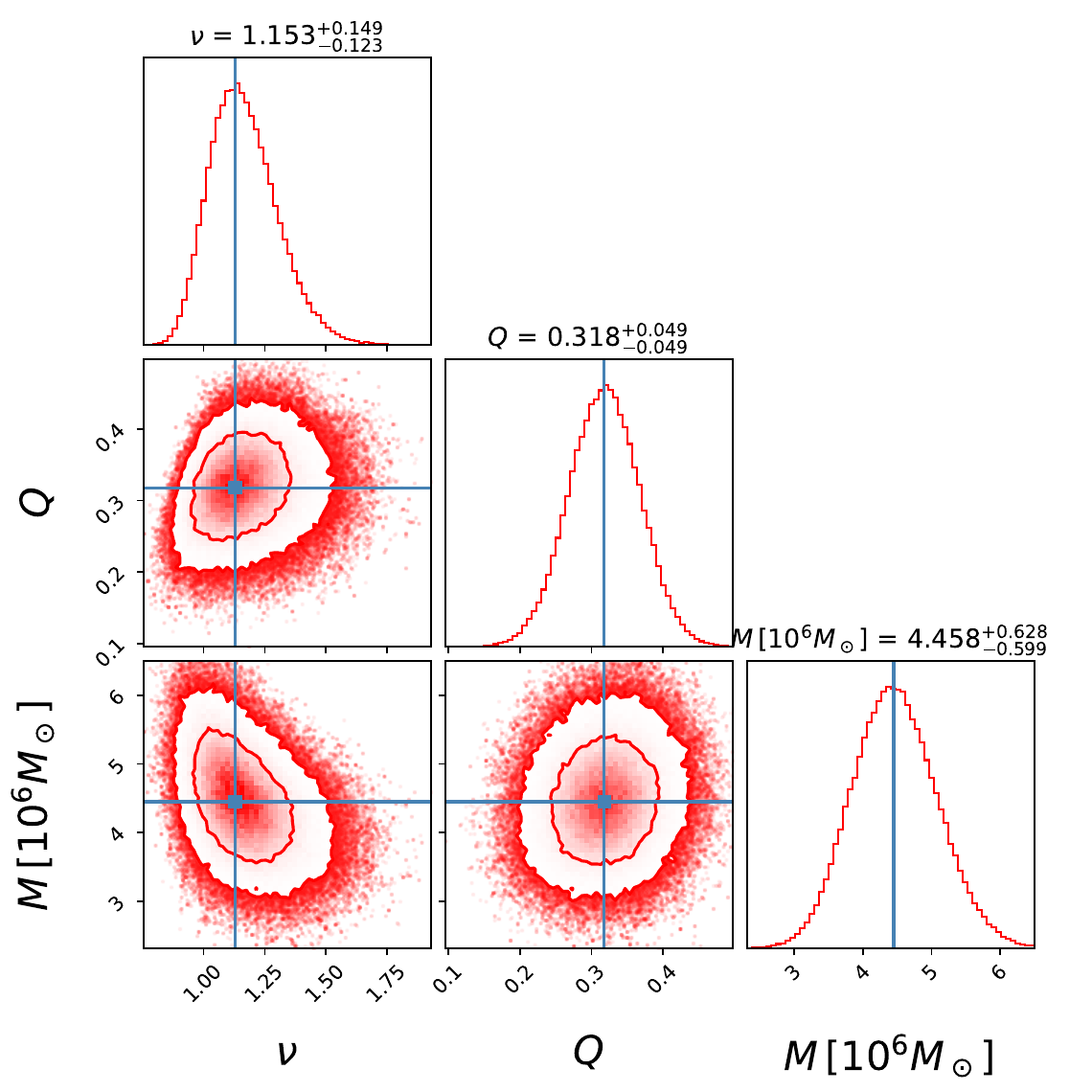}
\includegraphics[width=0.45\textwidth]{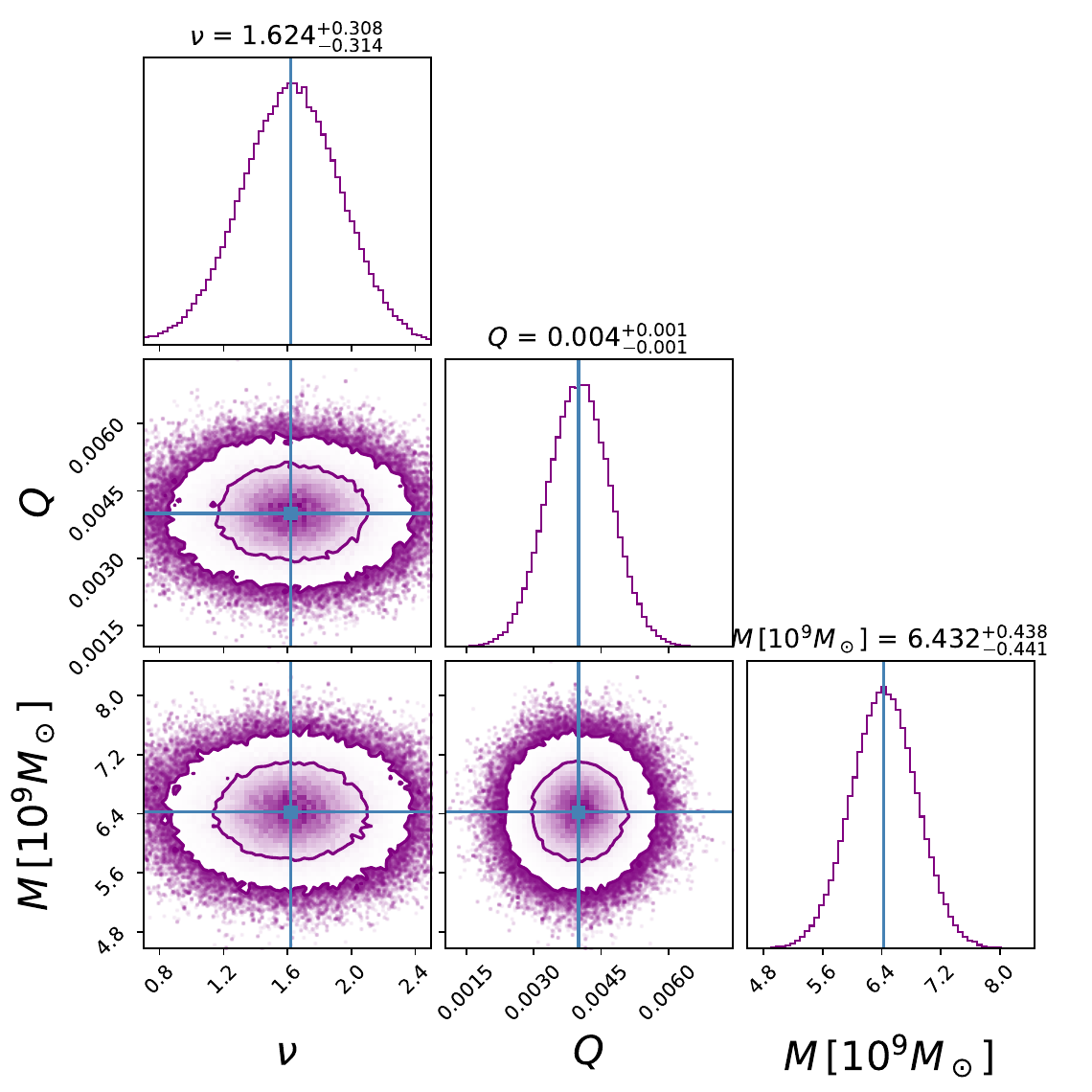}
\caption{
MCMC for Sgr A* from EHT (left) and from gravity (right). M87 bottom }
    \label{fig:MCMC}
\end{figure*}

Subsequently, we have employed MCMC analysis to constrain NED parameters $\nu$, $Q=\frac{q_m}{m}$ and BH mass $M$ and show visually in Fig.(\ref{fig:MCMC}). The results from the MCMC analysis  are summarized by the median value and the associated 68\% credible interval is given in Table (\ref{Table 3}). 

\begin{table}[ht!]
    \centering
    \resizebox{.47\textwidth}{!}{
    \begin{tabular}{|l|c|c|c|r|}
     \hline
        & $Sgr A^*$ from EHT & $Sgr A^*$ from Gravity & $M87^*$ from EHT \\
    \hline
      $M(\times 10^6M_\odot)$   & $4.229^{+0.616}_{-0.570}$    & $4.458^{+0.628}_{-0.599}$ & $6.432^{+0.432}_{-0.441}\times10^3$\\
      & & &\\
      $\nu$  & $1.195^{+0.239}_{-0.179}$   & $1.153^{+0.149}_{-0.123}$ & $1.624^{+0.308}_{-0.314}$\\
      & & &\\
      $Q$   & $0.253^{+0.049}_{-0.050}$   & $0.318^{+0.049}_{-0.049}$ & $0.004^{+0.001}_{-0.001}$\\
      
   \hline
    \end{tabular}}
    \caption{Best-fit generic regular BHs  parameters derived from angular shadow size.}
    \label{Table 3}
\end{table}

\section{GEODESICS EQUATIONS} \label{sec5}
To find the motion of the time-like particle we start with Hamilton-Jacobi equation (\cite{2025EPJC...85.1432U,2025PDU....4902022U,Uktamov:2024ckf}):
\begin{eqnarray}\label{eq.H-J equation}
    g^{\mu\nu}\frac{\partial S}{\partial x^{\mu}}\frac{\partial S}{\partial x^\nu}=-m^2\,,
\end{eqnarray}
where $S=-Et+L\phi+S_{\theta}(\theta)+S_r(r)$ is the Hamilton-Jacobi action.

Additionally, the Lagrangian $\mathcal{L}$ of the test particle in the vicinity of the generic regular BHs and corresponding conguate four-momentum $p$ can be written as:
\begin{subequations}\label{eq.lagrangian}
    \begin{align}
        &\mathcal{L}=\frac{1}{2}mg_{\mu\nu}u^{\mu}u^{\nu}\,,\\
        &p=\frac{\partial\mathcal{L}}{\partial\Dot{x}}=mg_{\mu\nu}u^{\nu}\,,
    \end{align}
\end{subequations}
where $u^{\mu}$ is the four-velocity of the particle.

\begin{figure*}[t]
\includegraphics[width=1\textwidth]{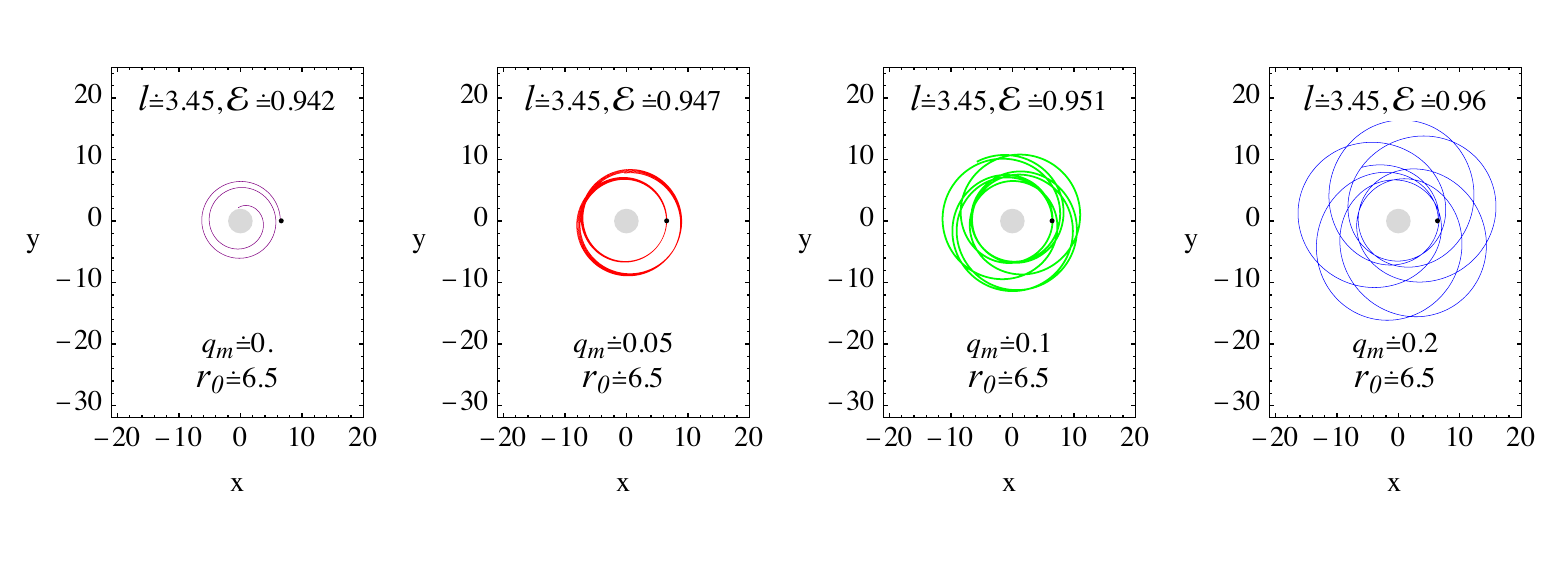}
\includegraphics[width=1\textwidth]{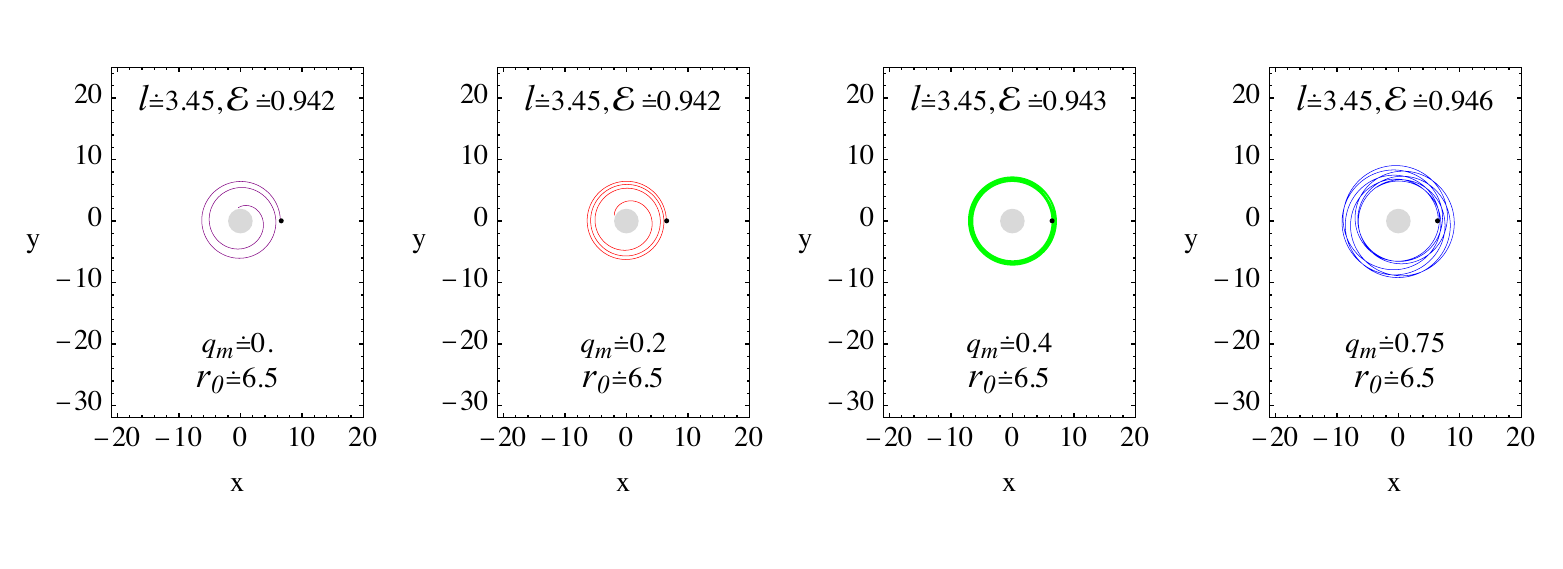}
\includegraphics[width=1\textwidth]{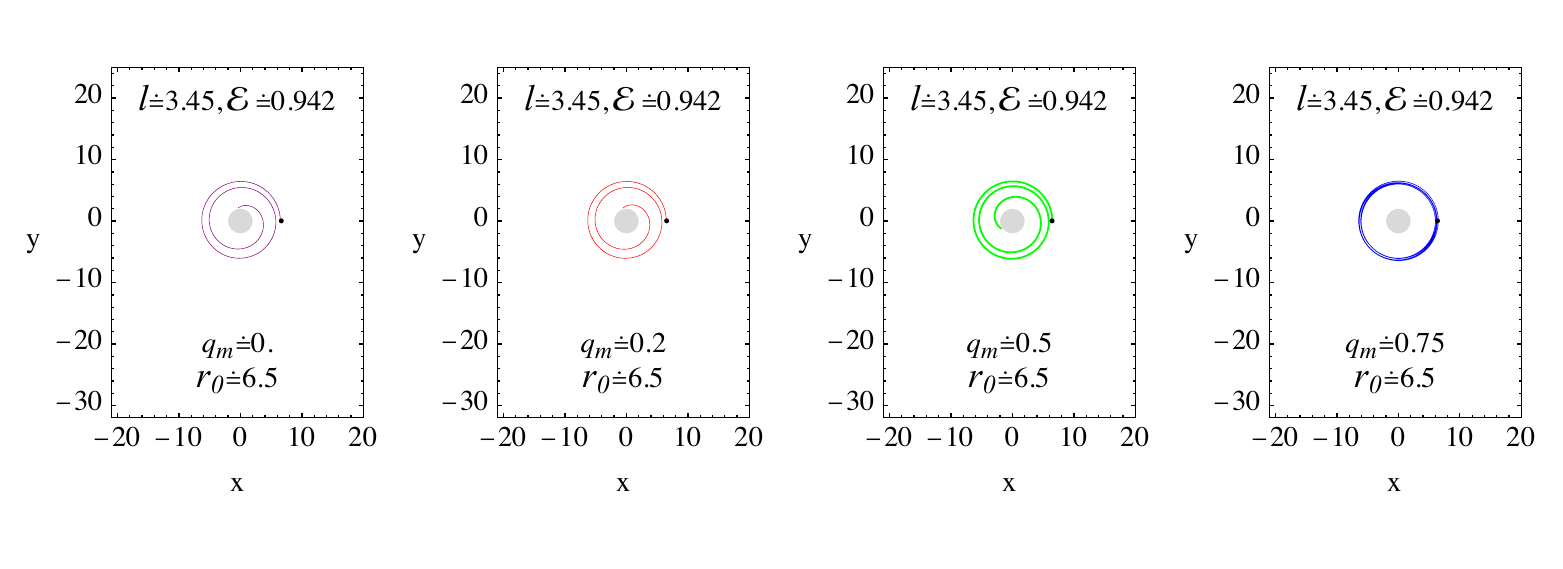}
\caption{The trajectory of the test particle orbiting around generic regular BHs (from top: first row for Maxwell solution, second row Bardeen solution, third row Hayward trajectory).}
    \label{fig:Trajectory}
\end{figure*}

Subsequently, we show trajectory of the test particle starting motion from the point $r_0$ in the vicinity of the generic regular BHs for different cases in Fig.(\ref{fig:Trajectory}) (\cite{Xamidov:2025hrj,Kolos:2023oii}). One can conclude from graphics in Fig.(\ref{fig:Trajectory}) that increasing the value of the magnetic charge $q_m$ causes growing specific energy $\mathcal{E}=\frac{E}{m}$ of the particles and make bound orbits. Also, it is clear from Fig.(\ref{fig:Trajectory}) that the effect of the magnetic charge $q_m$ on the trajectory of the test particle is very sensitive for Maxwellian solution.

If we restrict motion at the equatorial plane $\theta=\frac{\pi}{2}$,  $\Dot{\theta}=\Dot{\phi}=0$ (\cite{Ovgun:2025ehi}) as our metric (\ref{eq.the metric}) static and spherically symmetric, we will have expressions:
\begin{eqnarray}\label{eq.radial}
    \left(\frac{dr}{d\tau}\right)^2=\mathcal{E}^2-f(r)\,,\,\,\left(\frac{dr}{dt}\right)^2=(\frac{f(r)}{\mathcal{E}})^2\left[\mathcal{E}^2-f(r)\right],
\end{eqnarray}
where $\tau$ is the proper time.

\begin{figure*}[t]
\includegraphics[width=0.32\textwidth]{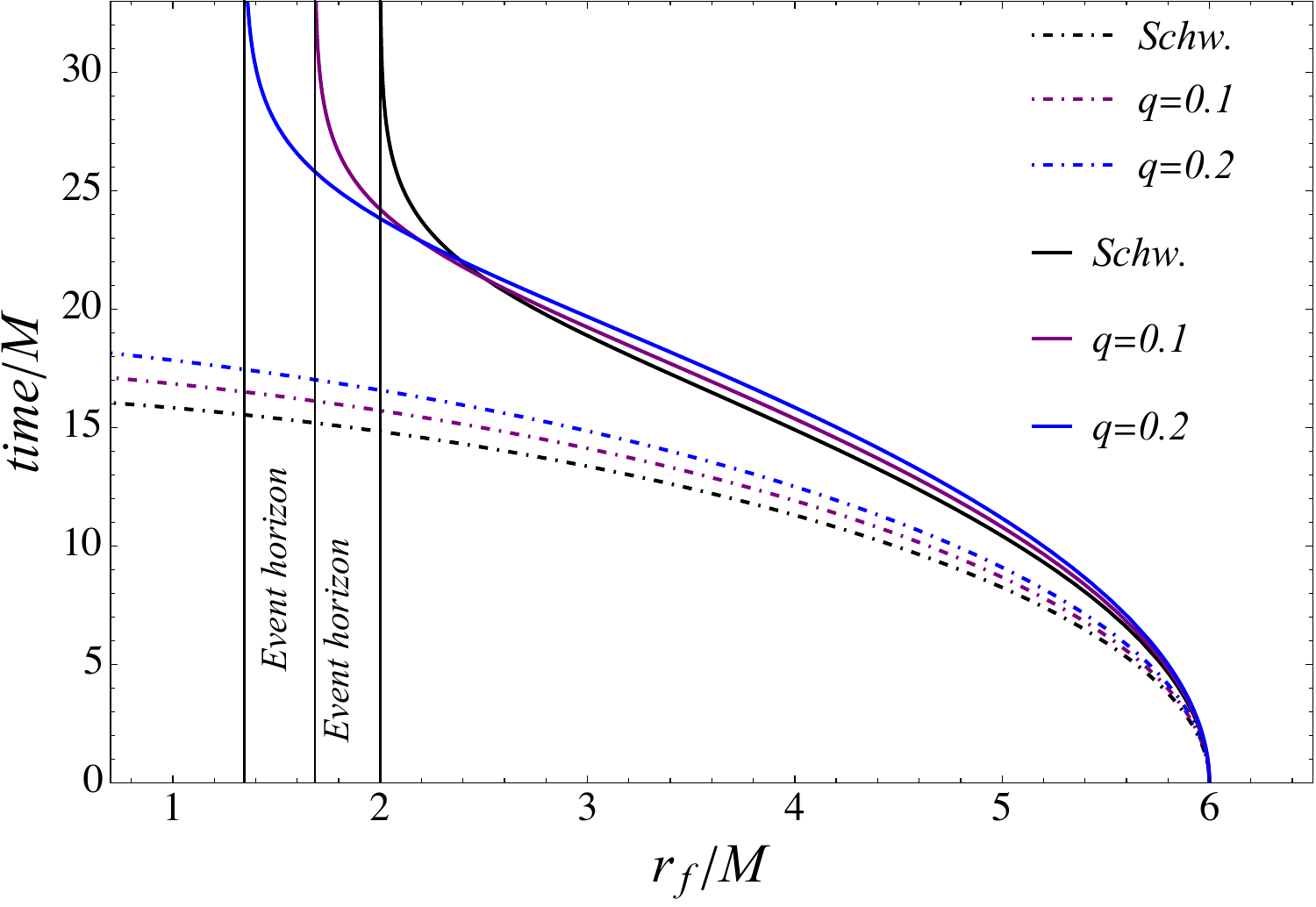}
\includegraphics[width=0.32\textwidth]{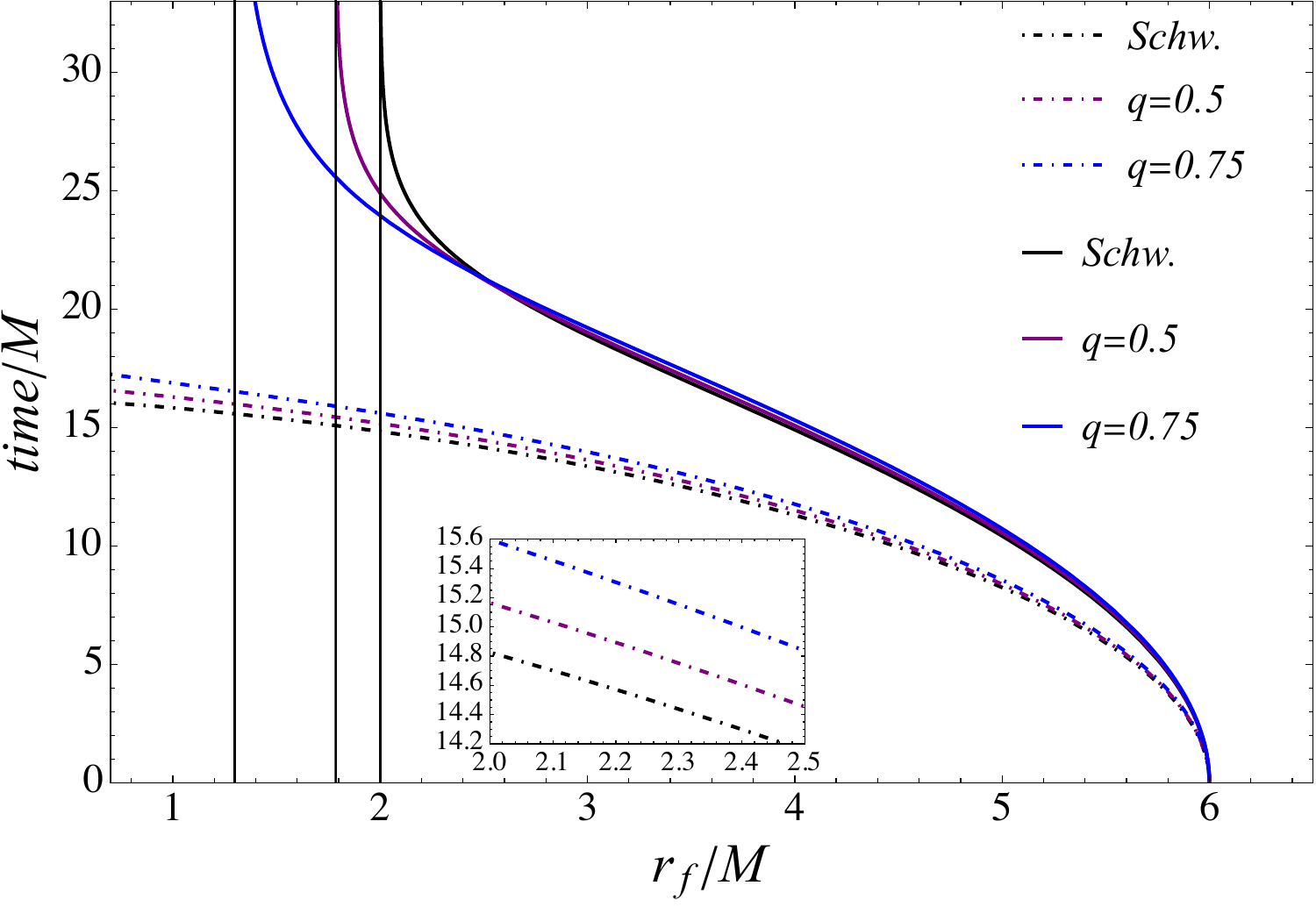}
\includegraphics[width=0.32\textwidth]{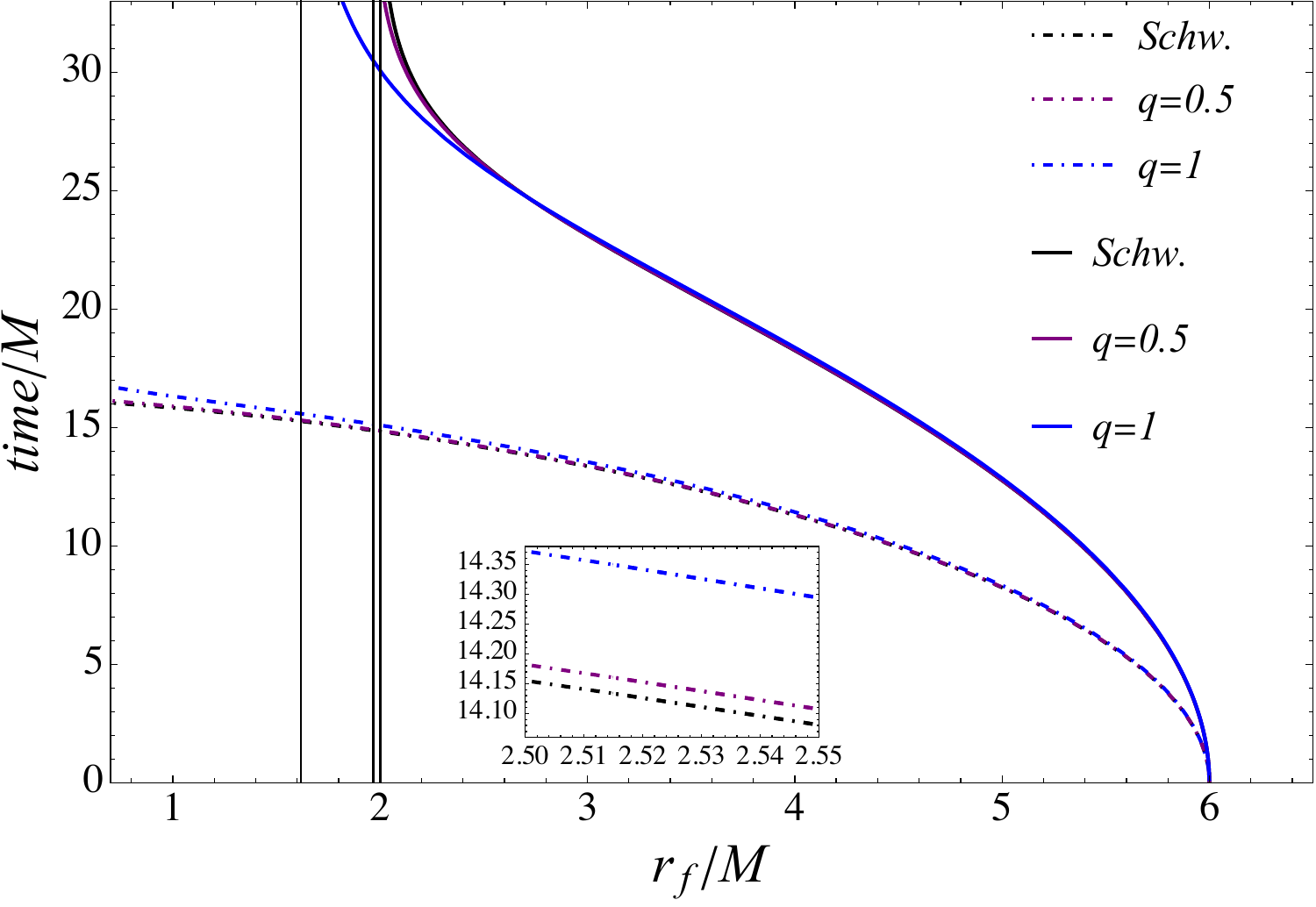}
\caption{The impact of the magnetic charge $q_m$ on the proper time $\tau$ (dotdashed lines) and coordinate time $t$ (solid lines) for Maxwell solution (the left panel), Bardeen solution (the middle panel) and Hayward solution (right panel). }
    \label{fig:Time}
\end{figure*}

Then for infalling particles from initial radius $r_i$ to final radius $r_f$ we can write:
\begin{eqnarray}\label{eq.time}
    \tau=-\int_{r_i}^{r_f}\frac{dr}{\sqrt{\mathcal{E}^2-f(r})}\,,\,\,\,t=-\int_{r_i}^{r_f}\frac{\mathcal{E}dr}{f(r)\sqrt{\mathcal{E}^2-f(r)}},
\end{eqnarray}
where we have expression for specific energy $\mathcal{E}^2=1-\frac{2Mr_i^{\mu-1}}{(r_i^\nu+q^\nu)^{\frac{\mu}{\nu}}}$ as $\Dot{r}=0$ when $r=r_i$. Subsequently, numerical integrating Eq.(\ref{eq.time}) paves a way to plot radial dependence of the proper time $\tau$ and coordinate time $t$ in Fig.(\ref{fig:Time}). One can see   \textit{frozen} star phenomena (\cite{2024PhRvD.110h4084U}) for coordinate time $t$ in Fig.(\ref{fig:Time}) also, it is clear from these plots (\ref{fig:Time}) that increasing the value of the magnetic charge $q_m$ leads enlarging falling time  of the particles.

Furthermore, hierarchical expansion $\overset{(\mathcal{H})}{\sim}$ near horizon enables us:
\begin{subequations}\label{eq.hier.}
    \begin{align}
        &f(r)\overset{(\mathcal{H})}{\sim}f'(r_g)x\left[1+\mathcal{O}(x)\right]\,,\\
        &f'(r)\overset{(\mathcal{H})}{\sim}f'(r_g)\left[1+\mathcal{O}(x)\right]\,,\\
        &f''(r)\overset{(\mathcal{H})}{\sim}f''(r_g)\left[1+\mathcal{O}(x)\right]\,,
    \end{align}
\end{subequations}
where $x=r-r_g$ is the shifted coordinate ().

Subsequently, using hierarchical expansion proper time $\tau$ and coordinate time $t$ can be calculated till first order in $x$ as:
\begin{subequations}\label{eq.time1}
    \begin{align}
        &\tau=-\frac{x}{\mathcal{E}}+\text{const.}+\mathcal{O}(x^2)\,,\\
        &t=-\frac{1}{f'(r_g)}\ln{x}-Cx+\text{const.}+\mathcal{O}(x^2)\,,
    \end{align}
\end{subequations}
in which:
\begin{eqnarray}
    C=\frac{1}{2}\left[\frac{1}{\mathcal{E}^2}-\frac{f''(r_g)}{[f'(r_g)]^2}\right]\,,
\end{eqnarray}
here we have used near horizon approximation:
\begin{subequations}
    \begin{align}
        &\frac{1}{f(r)}\approx\frac{1}{f'(r_g)x}\left[1-\frac{f''(r_g)}{2f'(r_g)}x\right]\,,\\
        &\frac{1}{\sqrt{\mathcal{E}^2-f(r)}}\approx\frac{1}{\mathcal{E}}+\frac{f'(r_g)x}{2\mathcal{E}^3}\,.
    \end{align}
\end{subequations}

\section{NEAR-HORIZON CONFORMAL QUANTUM MECHANICS EQUATION} \label{sec6}

In this section we compute the solution of the  Klein–Gordon equation in the generic regular BH (\ref{eq.the metric}) background so a massless Klein–Gordon field, when minimally coupled, obeys the following wave equation (\cite{Birrell:1982ix}):
\begin{eqnarray}\label{eq.Klein Gordon}
    \frac{1}{\sqrt{-g}}\partial_\alpha\left(\sqrt{-g}g^{\alpha\beta}\partial_\beta\Phi\right)=0\,,
\end{eqnarray}
where $g$ is the determinant of the metric (\ref{eq.the metric}).

For the analysis conducted here, the dominant contribution to the detector's response within the near-horizon region comes from the s-wave sector so Eq.(\ref{eq.Klein Gordon}) can be rewritten as:
\begin{eqnarray}\label{eq.Klein-Grodon2}
    -\frac{1}{f(r)}\partial_t^2\Phi+\partial_r\left[f(r)\partial_r\Phi\right]=0\,,
\end{eqnarray}
where radial equation reduces to the $(t\,,r)$ sector.
Then we can expand the field in modes as:
\begin{subequations}\label{eq.Field}
    \begin{align}
      &\Phi(r\,,t)=\sum\left[\hat{a}\phi(r\,,t)+\text{H.c.}\right]\,,\\
      &\phi(r\,,t)=\xi(r)u(r)e^{-i\nu_0t}\,,
    \end{align}
\end{subequations}
here H.c. is the hermitian conjugate, $\hat{a}$ is the field annihilation operator, $\nu_0$ is the mode frequency.

The function $\xi(r)$ is defined to eliminate the first-derivative term from the radial equation, thereby recasting it in the form of a Schr\"odinger-like form (\cite{Ovgun:2025ehi}):
\begin{eqnarray}\label{eq.xi}
    \xi(r)=\exp{\left(-\frac{1}{2}\int\frac{f'(r)}{f(r)}dr\right)}=\frac{1}{\sqrt{f(r)}}\,,
\end{eqnarray}
which paves a way to rewrite Eq.(\ref{eq.Klein-Grodon2}) as
\begin{subequations}\label{eq.u(r)}
    \begin{align}
      &u''(r)+V_D(r,\nu_0)u(r)=0\,,\\
      &V_D=\frac{1}{f^2(r)}\left[\nu_0^2+\frac{f'^2(r)}{4}\right]-\frac{f''(r)}{2f(r)}\,,
    \end{align}
\end{subequations}

Near the outer horizon $r=r_g$, define $x=r-r_g$ and expand $f(r)=f'(r_g)x+\mathcal{O}(x^2)$.
Then the effective potential behaves as
\begin{eqnarray}\label{eq.V_d}
V_D(r,\nu_0)\simeq \left[\frac{\nu_0^2}{f'^2(r_g)}+\frac{1}{4}\right]\frac{1}{x^2}
+\mathcal{O}\!\left(\frac{1}{x}\right),\end{eqnarray}
so that the conformal quantum mechanics (CQM)  equation becomes\begin{eqnarray}\label{eq.CQM equation}
u''(x)+\frac{\lambda_{\rm eff}}{x^2}u(x)=0,
\\
\lambda_{\rm eff}=\frac{1}{4}+\frac{\nu_0^2}{f'^2(r_g)}.
\end{eqnarray}

The solution to the Eq.(\ref{eq.CQM equation}) for the atomic outgoing radiation wave:
\begin{eqnarray}\label{eq.solut. CQM}
u(x)=\sqrt{x}x^{\frac{i\nu_0}{f'(r_g)}}\,.
\end{eqnarray}

Considering Eqs.(\ref{eq.xi},\ref{eq.solut. CQM}) Eq. (\ref{eq.Field}) can be rewritten as:
\begin{subequations}\label{eq.phi1}
    \begin{align}
        & \phi(r,t)=\frac{1}{\sqrt{f'(r_g)}}x^{i\Theta}e^{-i\nu_0t}\,,\\
        &\Theta=\frac{\nu_0}{f'(r_g)}\,.
    \end{align}
\end{subequations}

\section{ACCELERATION RADIATION FROM ATOMS FALLING INTO GENERIC REGULAR BLACK HOLES} \label{sec7}

In this section we consider the interaction between a scalar field and a two-level atom in geodesic motion, and derive the excitation probability for photon emission. Schematically, the setup consists of a central black hole and a mirror placed near the horizon. The mirror blocks Hawking radiation, so that any observed radiation arises solely from the atom–field interaction in the curved background. 

Using time-dependent perturbation theory, the excitation probability 
$P_{exc.}$ for the transition $|b\rangle\to|a\rangle$ with emission of a Killing-frequency $\nu$ quantum is:

\begin{eqnarray}\label{eq.Pexc}
        & P_{\rm exc}
    = \left|
        \int d\tau\,
        \langle 1_{\nu},a |
          V_{I}(\tau)
        | 0_{B},b\rangle
      \right|^{2}\,,
\end{eqnarray}
in which 
\begin{eqnarray}\label{eq.Vi}
    V_{I}(\tau)
    &=&g_c
      \bigl[
        \hat{a}_{\nu}\,\phi(r(\tau),t(\tau))
        + \hat{a}^{\dagger}_{\nu}\,\phi^{*}(r(\tau),t(\tau))
      \bigr]\\\nonumber
      &\times&\bigl(
        \sigma_{-} e^{-i\omega\tau}
        + \mathrm{H.c.}
      \bigr),
\end{eqnarray}
here $|1\rangle$ is the  one-photon mode with frequency $\nu$, $\sigma_{-}$ is the atomic lowering operator, $\omega$ is the atomic transition frequency, and $g_c$ is the atom-field coupling constant. 

Then Eq.(\ref{eq.Pexc}) can be rewritten in the form:
\begin{eqnarray}\label{eq.Pexc1}
 P_{\rm exc}
    = g_c^2
      \Bigl|
        \int d\tau\,
        \phi^{*}(r[\tau],t[\tau])\,e^{i\omega\tau}
      \Bigr|^{2}\,,    
\end{eqnarray}
as non zero component in Eq.(\ref{eq.Pexc}) comes from expression $\hat{a}^{\dagger}_{\nu}\sigma_{+}$.

Finally, considering Eqs.(\ref{eq.phi1}) and (\ref{eq.time1}) excitation probability $P_{exc}$ in Eq.(\ref{eq.Pexc1}) can be expressed as:
\begin{eqnarray}
    P_{\rm exc}
    = \frac{g_c^2}{\mathcal{E}^2}
      \Bigl|
        \int_0^{x_f} 
        x^{-2i\Theta}e^{-i\gamma x}
      dx\Bigr|^{2}\,,\\
      \end{eqnarray}
in which $x_f$ is the upper bound of the region where the near-horizon expansion is applicable and $\gamma$ is:

\begin{eqnarray}\nonumber
\gamma=\left(C\nu_0+\frac{\omega}{\mathcal{E}}\right)\,.
    \end{eqnarray}

In the case of the atom’s infall, the geometric optics regime assumes that $\omega>>\nu_0$ so that $\gamma>>\Theta$ as a consequence $x^{-2i\Theta}$, oscillates much more slowly than $e^{-i\gamma x}$ for larger $x$. Therefore the upper limit may be extended to infinity, and the integral can be evaluated via analytic continuation of the standard identities (\cite{Camblong2020}) :
\begin{subequations}\label{eq.form.}
    \begin{align}
        &\int_0^\infty x^{-i\sigma}e^{-ix}dx=e^{-\frac{\pi\sigma}{2}}e^{-\frac{i\pi}{2}}\Gamma(1-i\sigma)\,,\\
        &\Bigl|\Gamma(1-i\sigma)\Bigl|^2=\frac{\pi\sigma}{\sinh{(\pi\sigma)}}\,,
    \end{align}
\end{subequations}
which gives expression for $P_{exc.}$ as:
\begin{eqnarray}
    P_{exc.}=\frac{4\pi g_c^2\Theta}{\mathcal{E}^2\gamma^2}\frac{1}{e^{4\pi\Theta}-1}.
\end{eqnarray}

Fallowing, considering Eq.(\ref{eq.k and T}) and condition $\omega>>\nu_0$ $P_{exc.}$ can be expressed as in Planckian spectrum like form:
\begin{eqnarray}\label{eq.P2}
P_{exc.}\approx\frac{g_c^2\nu_0}{\omega^2T_H}\frac{1}{e^{\frac{\nu_0}{T_H}}-1}.
\end{eqnarray}

\begin{figure*}[t]
\includegraphics[width=0.45\textwidth]{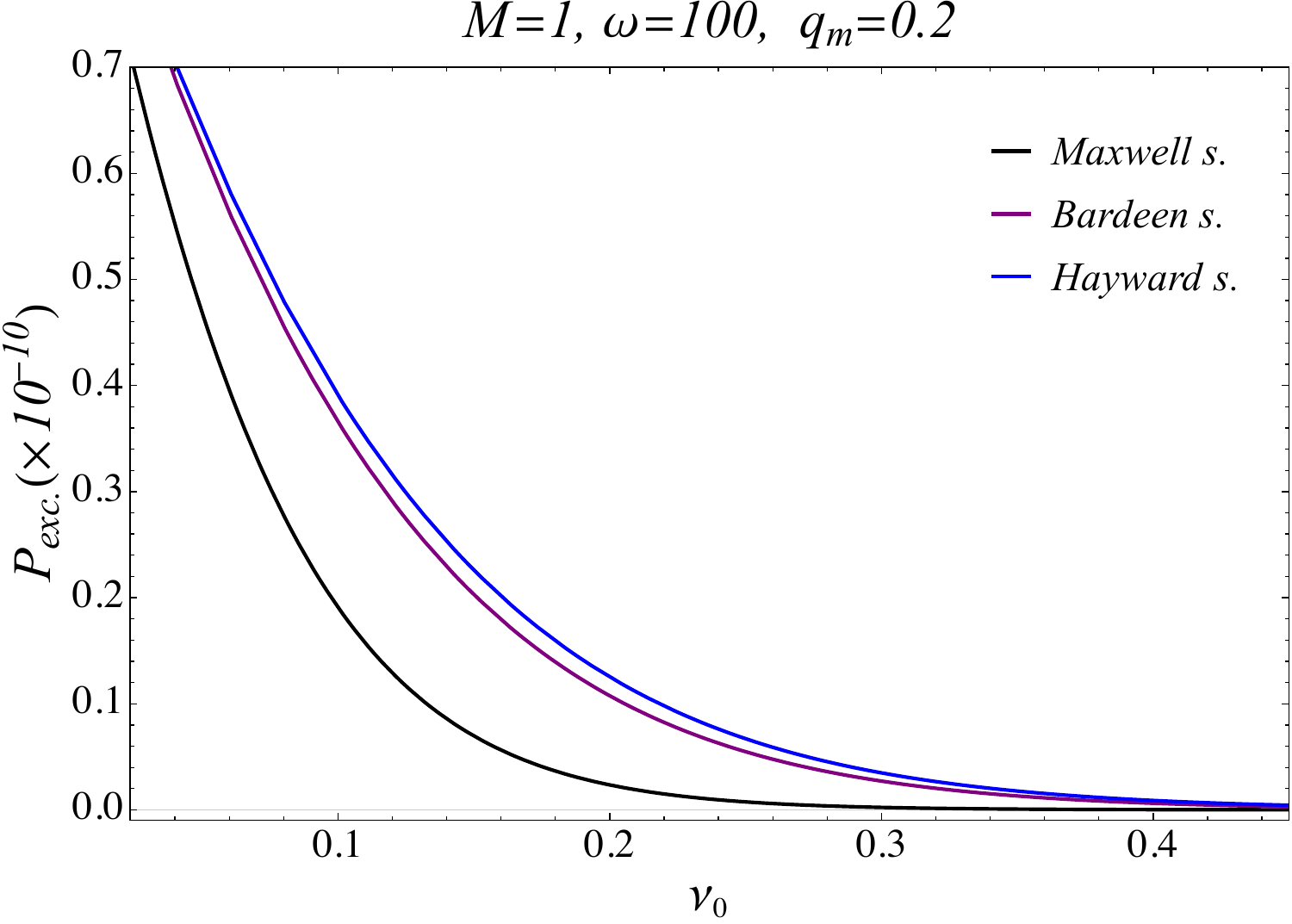}
\includegraphics[width=0.45\textwidth]{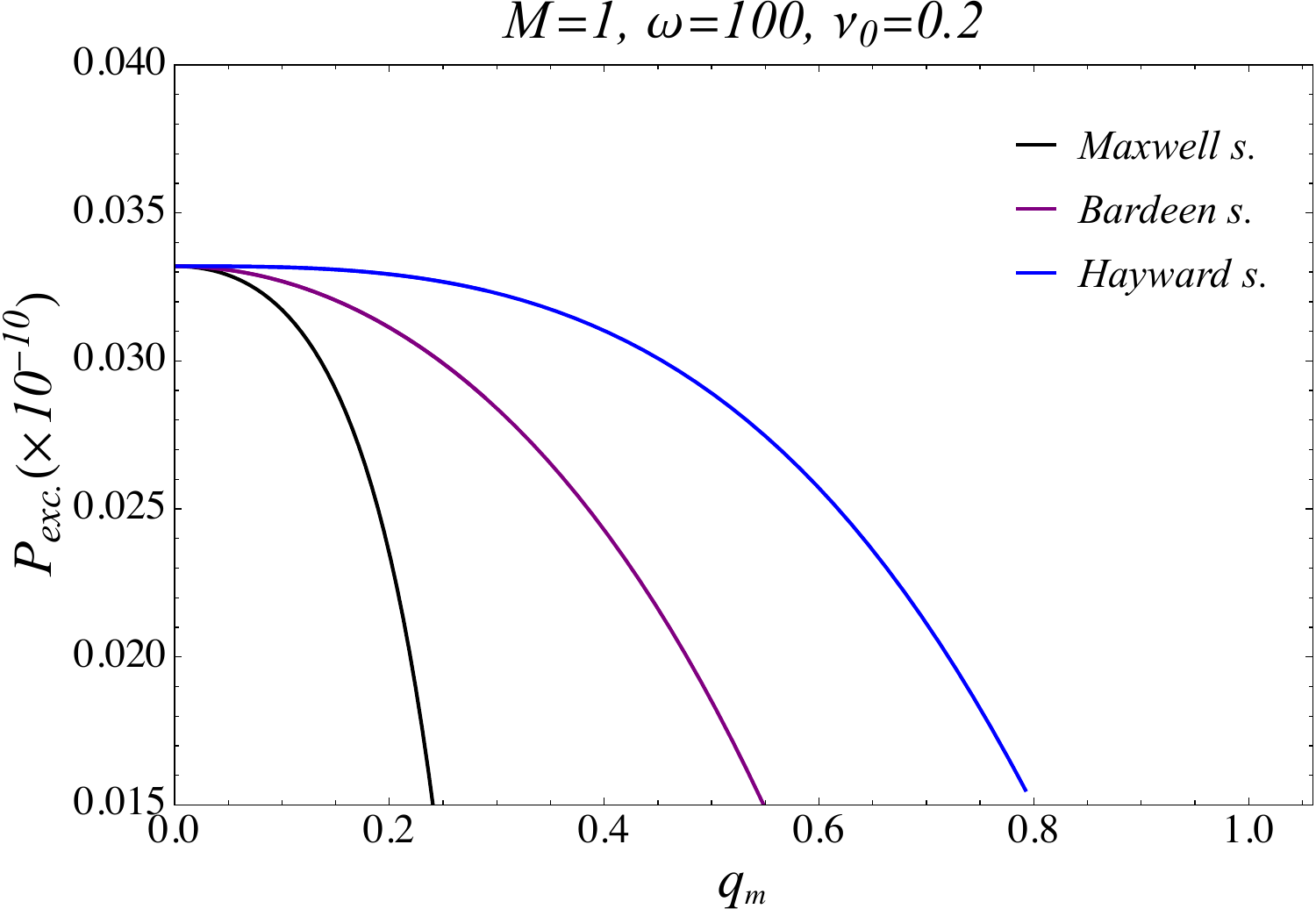}
\caption{
The dependence of the radiation intensity $P_{exc.}"$ on the mode frequencies $\nu_0$ (left panel) and on charge $q_m$ for different solutions. Here we take $g_c=10^{-3}$. }
    \label{fig:Probability}
\end{figure*}
\begin{figure*}[t]
\includegraphics[width=0.3\textwidth]{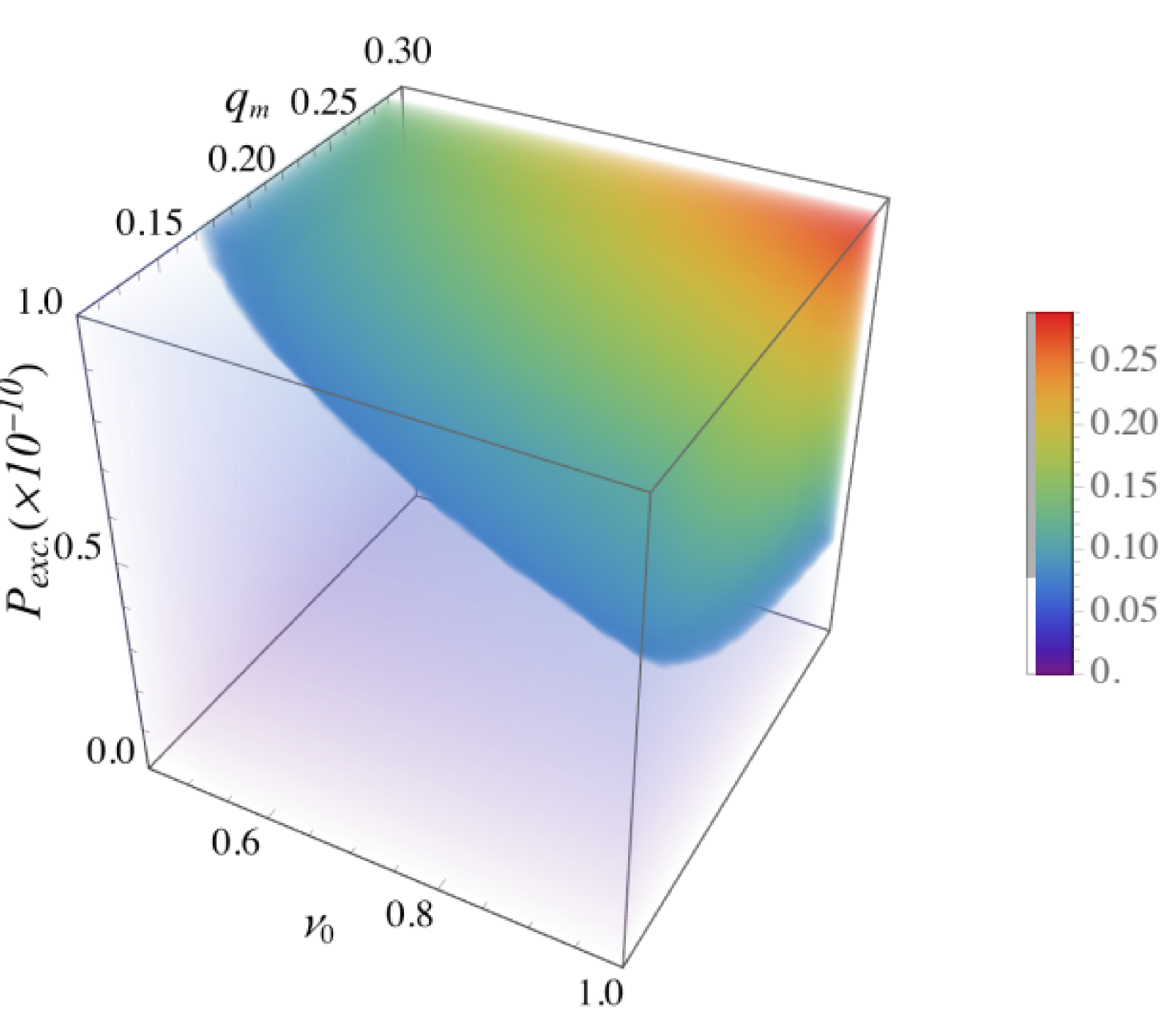}
\includegraphics[width=0.3\textwidth]{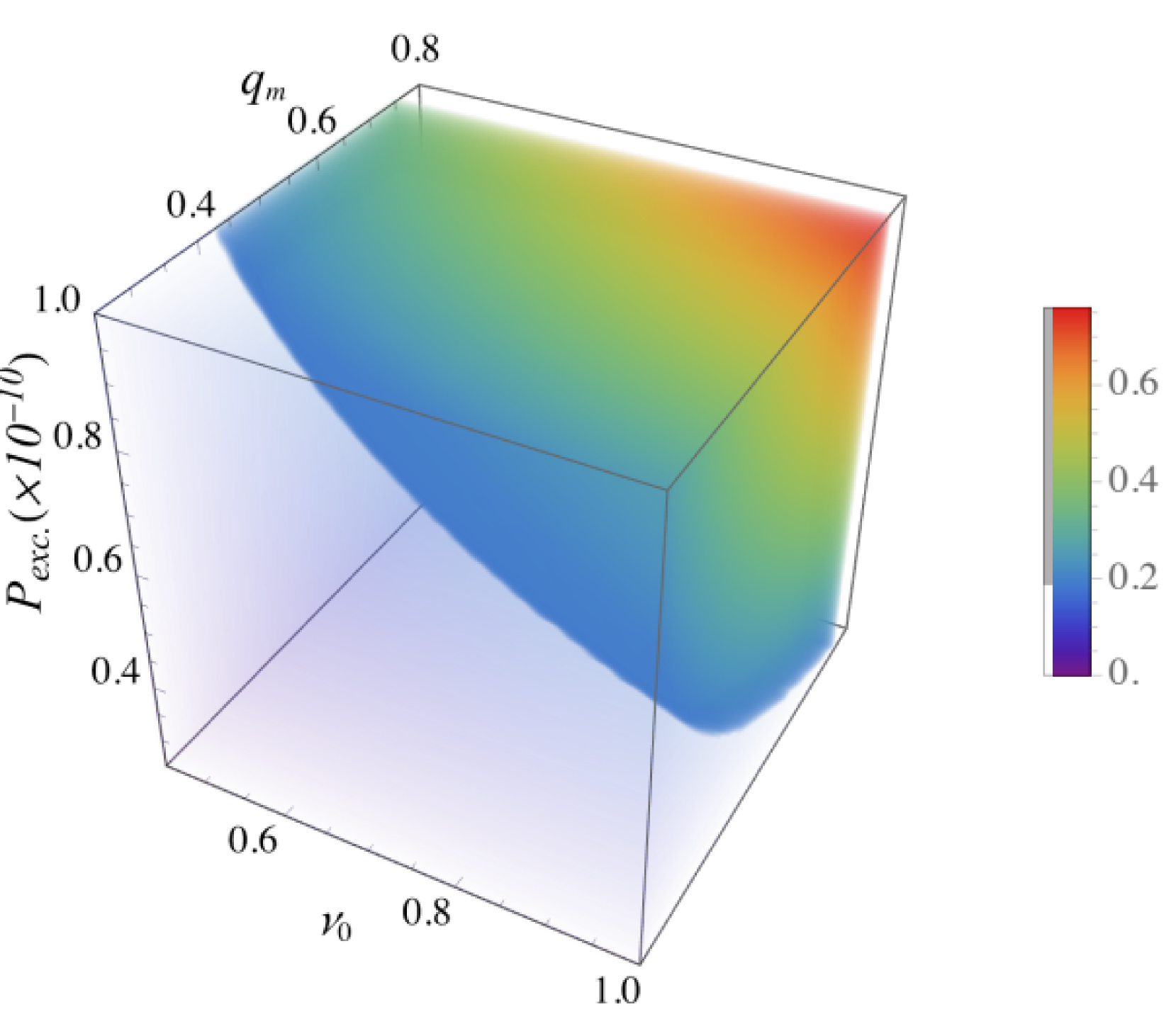}
\includegraphics[width=0.3\textwidth]{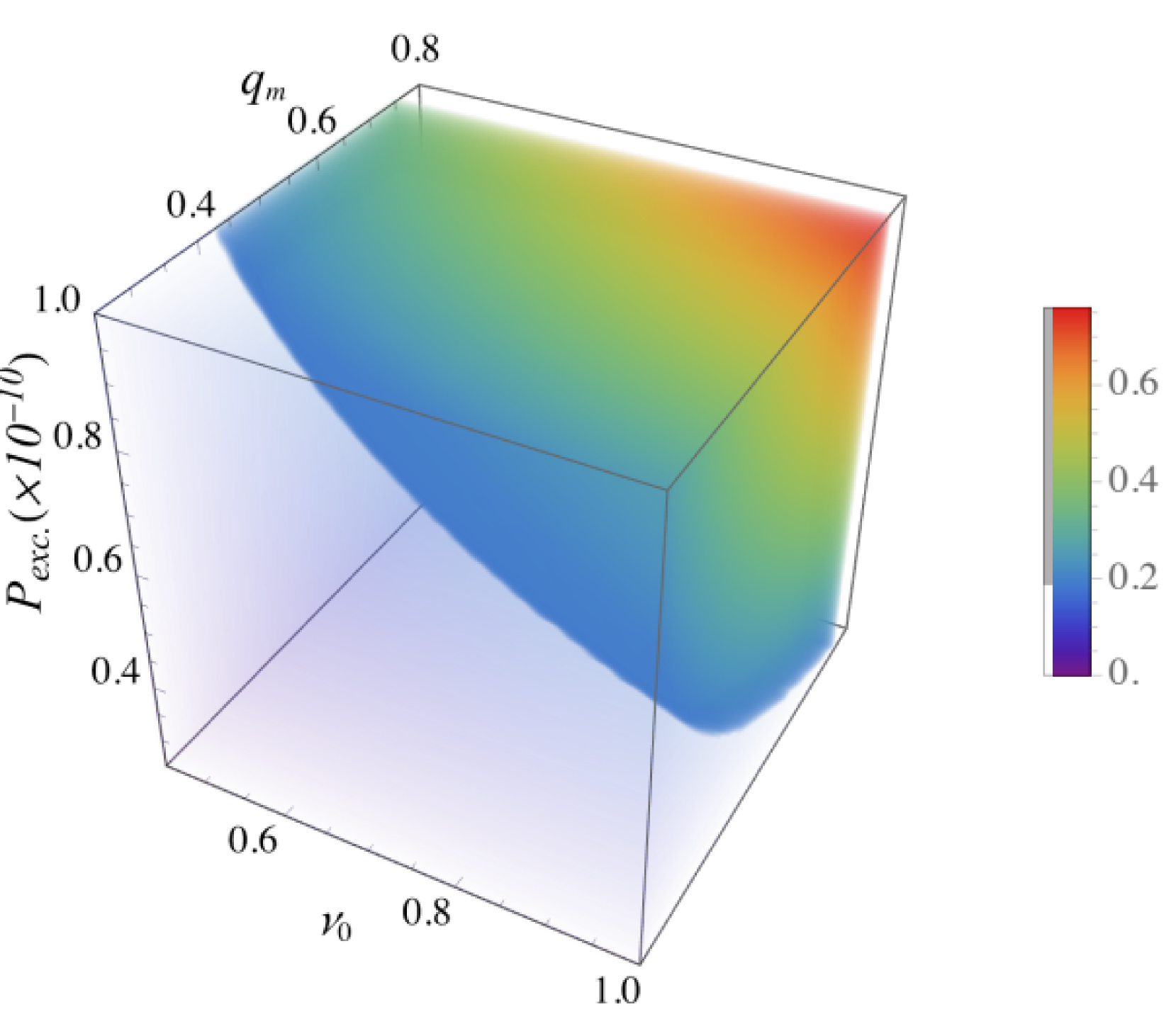}
\caption{
Radiation intensity $P_{exc.}$ plotted against mode frequencies $\nu_0$ and magnetic charge $q_m$ for Maxwell solution (left panel), Bardeen solution (middle panel), Hayward solution (right panel) in 3 dimensions.}
    \label{fig:Probability 3d}
\end{figure*}

Then we have plotted excitation probability $P_{exc.}$ as a function mode frequency $\nu_0$ and as a function magnetic charge $q_m$ in Fig.(\ref{fig:Probability}). Also,  HBAR's highest efficiency regions are presented through two and three-dimensional view in Fig. (\ref{fig:Probability 3d}).

\section{HBAR ENTROPY AND AREA LAW IN GENERIC REGULAR BLACK HOLES} \label{sec8}
Adopting the conventional laser-like description used in HBAR theory, the dynamical evolution of the mode density matrix $\rho(t)$ (in the interaction picture) is governed by a Lindblad master equation.
\begin{equation}
 \dot{\rho}
 =
 -i\,[H_{\rm eff},\rho]
 + \Gamma_{\rm em}\,\mathcal{D}[a^{\dagger}]\rho
 + \Gamma_{\rm abs}\,\mathcal{D}[a]\rho ,
 \label{eq:master}
\end{equation}
where $H_{\rm eff}=\nu_0\, a^{\dagger}a$ is the effective Hamiltonian, $\mathcal{D}[O]\rho$ is the dissipator (\cite{Ovgun:2025ehi}):
\begin{equation}
 \mathcal{D}[O]\rho
 = O\,\rho\,O^{\dagger}
 - \frac{1}{2}\left\{O^{\dagger}O,\rho\right\}\,.
\end{equation}
Also, considering as two-level atoms with gap $\omega$ are injected into the cavity at an average flux $J$ (number of atoms per unit Killing time), the emission $\Gamma_{em}$ and absorption $\Gamma_{abs}$ rates can be expressed as:
\begin{eqnarray}
    \Gamma_{\rm em} = \mathcal{J}\,P_{\rm exc},
 \,\,
 \Gamma_{\rm abs} = \mathcal{J}\,P_{\rm abs}\,,
\end{eqnarray}
or using Eq.(\ref{eq.P2}):
\begin{eqnarray}\label{eq.Gamma}
 \frac{\Gamma_{em}}{\Gamma_{abs}}=e^{-\frac{\nu_0}{T_H}}\,.   
\end{eqnarray}
In the Fock basis, the diagonal element of the density matrix $\rho_{n,n}(t)$ satisfy:
\begin{eqnarray}
\dot{\rho}_{n,n}
&=& \Gamma_{\rm em}\Bigl[n\,\rho_{n-1,n-1}-(n+1)\rho_{n,n}\Bigr]
\\\nonumber
&+& \Gamma_{\rm abs}\Bigl[(n+1)\rho_{n+1,n+1}-n\,\rho_{n,n}\Bigr]\,.
\label{eq:rate_rho}
\end{eqnarray}
Subsequently, employing stationary condition $\dot{\rho}_{n,n}=0$ and Eq.(\ref{eq.Gamma}), we will have:
\begin{eqnarray}
    \frac{\rho^{(\mathrm{st})}_{n,n}}{\rho^{(\mathrm{st})}_{n-1,n-1}}
=\frac{\Gamma_{\rm em}}{\Gamma_{\rm abs}}=e^{-\frac{\nu_0}{T_H}}\,,
\end{eqnarray}
which yields to the equation:
\begin{eqnarray}\label{eq.n1}
    \rho^{(\mathrm{st})}_{n,n}
=\bigl(1-e^{-\frac{\nu_0}{T_H}}\bigr)\,e^{-\frac{n\nu_0}{T_H}},
\,
\bar n_\nu
=\sum_n n\,\rho^{(\mathrm{st})}_{n,n}
=\frac{1}{e^{\frac{\nu_0}{T_H}}-1},
\end{eqnarray}
which is Bose-Einstein steady state.

\subsection{HBAR energy and entropy flux} \label{ssec8.1}

The energy contained within the horizon-brightened acceleration radiation (HBAR) field is
\begin{equation}
 E_{P}(t)
 = \sum_{\nu_0} \nu_0\,\bar{n}_{\nu_0}(t),
 \qquad
 \dot{E}_{P}
 = \sum_{\nu_0} \nu_0\,\dot{\bar{n}}_{\nu_0},
 \label{eq:EP_def}
\end{equation}
where in the continuum limit, the summation over $\nu_0$ can be expressed as an integral.

From the master Eq. ~\eqref{eq:master} enables to find the mode occupation number
\begin{equation}\label{eq.n}
\dot{\bar n}_{\nu_0}
=\Gamma_{\rm em}\,(\bar n_{\nu_0}+1)-\Gamma_{\rm abs}\,\bar n_{\nu_0}
=\Gamma_{\rm em}-(\Gamma_{\rm abs}-\Gamma_{\rm em})\,\bar n_{\nu_0}.
\end{equation}
Eq.(\ref{eq.n}) reduces as:  
\begin{equation}
\dot{\bar{n}}_{\nu_0}\simeq \Gamma_{\rm em}
=\mathcal{J}\,P_{\rm exc}\,,
\label{eq:nbar_dot_approx}
\end{equation}

in the  dilute regime $\bar n_{\nu_0}\ll1$ and we use below to estimate the initial HBAR energy/entropy production rates.

 Substituting Eqs. \eqref{eq:EP_def} and~\eqref{eq:nbar_dot_approx} into Eq. ~\eqref{eq.P2} yields the HBAR energy flux in the Generic regular black hole spacetimes in the form
\begin{equation}
 \dot{E}_{P}(g)
 \simeq
 \frac{ g_c^{2}\mathcal{J}}{\omega^{2}}
 \sum_{\nu_0}
 \frac{\nu_0^{2}}{T_H}\,
 \frac{1}{e^{\frac{\nu_0}{T_H}}-1}.
 \label{eq:EP_flux}
\end{equation}
The dependence on the NED $\nu$ enters exclusively through the Hawking temperature $T_{H}^{(\mathrm{B})}(q_m,\nu)$.

The von Neumann entropy of the radiation field is
\begin{equation}
 S_{\rho}(t)
 = -\sum_{n,\nu} \rho_{n,n}(t)\,\ln\rho_{n,n}(t),
 \label{eq:Srho_def}
\end{equation}
and its rate of change resulting from photon emission is
\begin{equation}
 \dot{S}_{\rho}
 = -\sum_{n,\nu} \dot{\rho}_{n,n}\,\ln \rho_{n,n}^{(\mathrm{st})}.
 \label{eq:Srho_dot_def}
\end{equation}
Employing  the master equation (\ref{eq:master}) and the steady-state distribution~\eqref{eq.n1}, one finds Schwarzschild-like relationship between HBAR entropy and energy flux:
\begin{equation}
 \dot{S}_{\rho}
 = \frac{\dot{E}_{P}}{T_H}=\frac{4\pi r_g\dot{E}_{P}\left(r_g^\nu+q_m^\nu\right)}{r_g^\nu-2q_m^\nu}\,,
 \label{eq:HBAR_first_law}
\end{equation}
which is the HBAR analogue of the Clausius relation.

\section{WIEN'S DISPLACEMENT LAW IN THE
GENERIC REGULAR BLACK HOLES} \label{sec9}
The thermal nature of the HBAR excitation probability (\ref{eq.P2}) leads to a Wien-type displacement law within the generic regular black holes. The spectral weight per unit wavelength, denoted by $P_\lambda(\lambda)$, is defined according to:
\begin{eqnarray}\label{eq.P3}
    P_{\lambda}(\lambda)d\lambda&=&P_{\nu_0}(\nu_0)d\nu_0\,,\,\,\,\,\nu_0=\frac{1}{\lambda},\\\nonumber
    d\nu_0&=&-\frac{1}{\lambda^2}d\lambda\,,
\end{eqnarray}
here we write $P_\lambda(\lambda)$ in terms of the wavelength $\lambda$:
\begin{equation}
 \mathcal{P}_{\lambda}(\lambda)
 \propto
 \frac{1}{\lambda^{3}}\,
 \frac{1}{\exp\!\left(\dfrac{1}{\lambda T_{H}}\right)-1}.
 \label{eq:Plambda_spectrum}
\end{equation}
It is clear from Eq.(\ref{eq:Plambda_spectrum}) that all NED information is encoded in single Hawking temperature $T_H$ and the peak of the spectrum can be calculated by the condition:
\begin{equation}
 \frac{d\mathcal{P}_{\lambda}(\lambda;g)}{d\lambda}\bigg|_{\lambda=\lambda_{\rm crit}} = 0.
 \label{eq:extremum_condition}
\end{equation}
then after introducing new variable as $x=\frac{1}{\lambda T_H}$ we will have expression:
\begin{equation}
 \frac{d}{dx}\left[
   \frac{x^{3}}{e^{x}-1}
 \right]_{x=x_{\rm crit}} = 0\,,
\end{equation}
which gives equation:
\begin{equation}
 1 - e^{-x_{\rm crit}} = \frac{x_{\rm crit}}{3},
 \label{eq:Wien_equation}
\end{equation}
solving Eq.(\ref{eq:Wien_equation}) numerically yields to the unique positive solution $x_{\rm crit}\simeq 2.82144$. Then restoring constants $\hbar$ and $k_{B}$ paves a way to write:
\begin{eqnarray}
\lambda_{\rm crit}^{(\mathrm{B})}
 &\simeq&
 \frac{2\pi}{x_{\rm crit}}
 \frac{\hbar}{k_B\,T_H}
 \approx 2.23\,\frac{\hbar}{k_B\,T_H}=\\\nonumber
&=&2.23\,\frac{2hr_g\left(r_g^\nu+q_m^\nu\right)}{k_B\,\left(r_g^\nu-2q_m^\nu\right)}.
 \label{eq:Wien_Bardeen_units}
\end{eqnarray}

\begin{figure*}[t]
\includegraphics[width=0.5\textwidth]{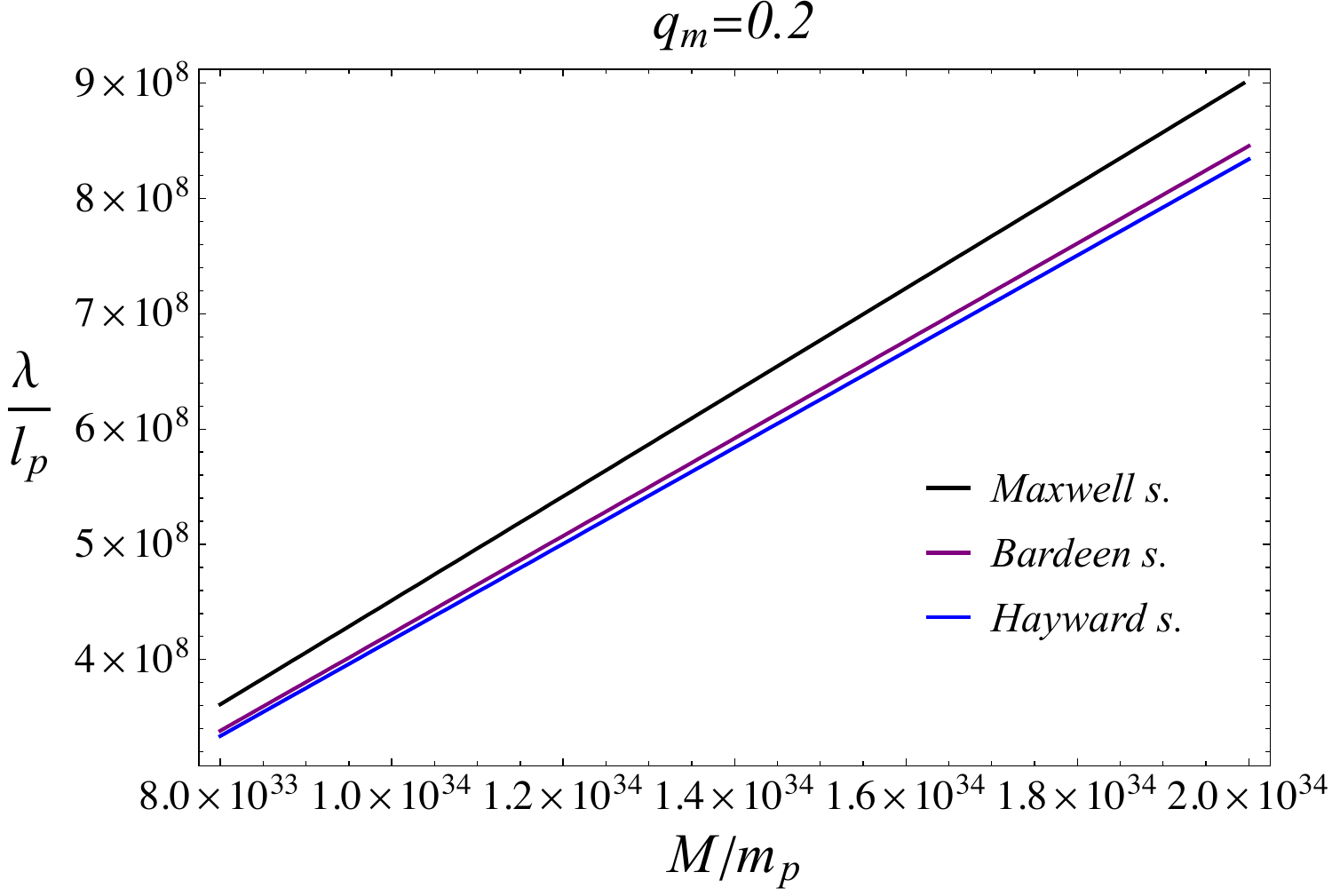}
\caption{
The dependence  of the wavelength ($\lambda/l_p$) on the  mass ($M/m_p$) of the black hole from the Wien’s displacement law . Here $l_p$ and $m_p$ indicate the Planck length and Planck mass respectively. }
    \label{fig:Wien}
\end{figure*}
Then we have shown variation of the peak Hawking wavelength with $M/m_p$ deived from Wien's displacement law. in Fig.(\ref{fig:Wien}).

\section{CONCLUSION} \label{sec10}

In this work we study two complementary strong-gravity probes for generic regular black holes supported by NED:    (i) optical signatures controlled by the photon region (shadow formation), and (ii) an operational near-horizon quantum channel, namely horizon–brightened acceleration radiation (HBAR) generated by infalling two-level atoms interacting with field modes close to the horizon.

On the classical side, we analyzed a broad parameterized family of static, spherically symmetric regular geometries that includes the well-known, Bardeen-like, and Hayward-like limits. We mapped how the horizon structure depends on the NED parameters and magnetic charge, identified the extremal boundary separating black-hole and no-horizon configurations, and studied the corresponding behavior of the Hawking temperature, which generically rises from the weak-charge regime, reaches a maximum, and then decreases to zero as extremality is approached.

We then derived the photon-sphere condition and determined the shadow radius for distant observers. Across the representative NED branches explored, increasing the dimensionless magnetic charge produces a systematic decrease of the photon-sphere radius and therefore a shrinking of the black-hole shadow, consistent with our numerical scans and shadow images. We also provided controlled analytic approximations in the small-charge regime, clarifying how departures from the Schwarzschild shadow scale with the NED parameters.

To connect these predictions with observations, we performed a Bayesian parameter inference using Markov Chain Monte Carlo sampling, employing angular shadow-size constraints from EHT and GRAVITY for Sgr A* and from EHT for M87*. The resulting posterior distributions constrain the allowed region of NED parameter space and quantify correlations between the inferred black-hole mass and the nonlinear-electrodynamic parameters. Within the adopted modeling assumptions, current horizon-scale data already place meaningful bounds on the dimensionless charge ratio and favor NED indices of order unity in the fits reported.

On the quantum side, we developed the near-horizon framework required for HBAR. By reducing the relevant field dynamics to the dominant near-horizon sector, we showed that the response of the infalling atomic detectors is governed by a conformal near-horizon structure and yields a thermal excitation spectrum controlled by the horizon temperature. We then adopted a Lindblad master-equation description for the radiation field, demonstrated the emergence of a thermal steady state, and established a consistent entropy–energy flux relation for HBAR that mirrors the thermodynamic structure expected of horizon-associated radiation. Finally, we formulated a Wien-type displacement law for the HBAR spectrum, relating the peak wavelength directly to horizon thermodynamic data, thereby providing an additional observable handle on NED regular black holes through near-horizon radiative physics.

Several extensions naturally follow. A more complete optical treatment can incorporate the possibility of NED-induced effective optical propagation effects where appropriate and explore finite-distance observables beyond the asymptotic shadow size. On the HBAR side, one may go beyond the dominant sector and dilute limit, include additional mode structure, and assess the impact of greybody factors and backreaction when aiming at quantitative flux forecasts. Extending the analysis to rotating regular black holes would be particularly valuable, as it would allow simultaneous confrontation of spin-dependent shadow observables with the operational HBAR channel. Overall, our results show that regularity-motivated NED black holes admit a coherent joint description in which shadow constraints and HBAR thermodynamics are controlled by the same horizon-scale data, providing a concrete pathway to test nonsingular strong-field models against present and future horizon-scale measurements.

\acknowledgments
 A. \"O. and R. P. would like to acknowledge networking support of the COST Action CA21106 - COSMIC WISPers in the Dark Universe: Theory, astrophysics and experiments (CosmicWISPers), the COST Action CA22113 - Fundamental challenges in theoretical physics (THEORY-CHALLENGES), the COST Action CA21136 - Addressing observational tensions in cosmology with systematics and fundamental physics (CosmoVerse), the COST Action CA23130 - Bridging high and low energies in search of quantum gravity (BridgeQG), and the COST Action CA23115 - Relativistic Quantum Information (RQI) funded by COST (European Cooperation in Science and Technology). A. \"O. also thanks to EMU, TUBITAK, ULAKBIM (Turkiye) and SCOAP3 (Switzerland) for their support.

\appendix

\allowdisplaybreaks


\bibliography{prd/main}    
\end{document}